\documentclass[onecolumn]{aastex7}
\usepackage{graphicx}
\usepackage{subfigure}

\begin{document}

\title{Searching for Accreting Compact Object Binaries in SRG/eROSITA eRASS1}
\shorttitle{Accreting Compact Object Binaries in eRASS1}
\shortauthors{Liu et al.}

\author[0000-0002-2912-095X]{Hao-Bin Liu}
\affiliation{Department of Astronomy, Xiamen University, Xiamen, Fujian 361005, People's Republic of China}
\email{liuhaobin@stu.xmu.edu.cn}

\author[0000-0003-3137-1851]{Wei-Min Gu}
\affiliation{Department of Astronomy, Xiamen University, Xiamen, Fujian 361005, People's Republic of China}
\email[show]{guwm@xmu.edu.cn}

\author[0009-0002-1645-7569]{Yongqi Lu}
\affiliation{Department of Astronomy, Xiamen University, Xiamen, Fujian 361005, People's Republic of China}
\email{luyongqi@stu.xmu.edu.cn}

\author[0000-0002-2941-6734]{Teng Liu}
\affiliation{School of Astronomy and Space Science, University of Science and Technology of China, Hefei, Anhui 230026, People's Republic of China}
\email{liuteng@ustc.edu.cn}

\author[0000-0002-7420-6744]{Jin-Zhong Liu}
\affiliation{Xinjiang Astronomical Observatory, Chinese Academy of Sciences, Urumqi, Xinjiang 830011, People's Republic of China}
\email{liujinzh@xao.ac.cn}

\begin{abstract}

Compact object binaries with accreting white dwarfs, neutron stars, or black holes are crucial for understanding accretion physics. In this study, we identify accreting compact object binary candidates in the SRG/eROSITA eRASS1 by combining their X-ray fluxes with Gaia photometry and ZTF time-domain observations. Candidates are selected based on their location in the "X-ray Main Sequence", a diagram incorporating their X-ray-to-optical flux ratios and optical colors, which suggest accretion-driven X-ray emission. We identify 22 candidates in eRASS1 catalog using a three-step selection process: (1) cross-matching to a unique Gaia optical counterpart within a 10$''$ radius; (2) requiring X-ray-to-optical flux ratios exceeding the "X-ray Main Sequence"; and (3) detecting short-period variability in ZTF time-domain photometry. The resulting 22 candidates, including two previously confirmed compact object binaries, represent promising candidates for spectroscopic follow-up to confirm their accreting nature. Our results demonstrate the effectiveness of combining X-ray-to-optical flux ratios and optical colors jointly with time-domain photometry to uncover accreting compact object binaries. The approach is scalable and adaptable to future multi-wavelength sky surveys, offering a promising path toward a more complete census of compact object binaries in the Galaxy.

\end{abstract}

\section{Introduction} \label{sec:intro}
Compact object binaries, pairing a compact object (white dwarf, neutron star, or black hole) with a companion star, are crucial for probing extreme astrophysical processes, including mass accretion and binary evolution \citep{2013Abramowicz, 2014LRR....17....3P, 2023hxga.book..129B}. The detection of compact object binary integrates multiple observational techniques including X-ray observations, which aim to detect those undergoing mass transfer, such as X-ray binaries and cataclysmic variables (CVs) \citep{2006Remillard, 1995Warner, 1997A&A...327..602V}. X-ray binaries have been the cornerstone of black hole searches, with discoveries relying on X-ray outbursts reaching luminosities of $10^{35}$--$10^{38}$~erg~s$^{-1}$ \citep{2016A&A...587A..61C}. However, outburst-based searches lack the sensitivity to identify the numerous Galactic quiescent X-ray binaries (QXBs) with X-ray luminosities of $10^{30}$--$10^{33}$~erg~s$^{-1}$ \citep{2000A&A...360..575L}. Novel approaches are therefore essential to uncover these faint systems and expand the census of compact object binaries.

Studies have employed X-ray observations to search for QXBs in specific regions \citep{1980PASP...92..691H, 2003ApJ...598..501H}. The Chandra Galactic Bulge Survey, led by \cite{2011AAS...21714424H}, used optical spectroscopy to characterize X-ray sources, while the Swift Bulge Survey, led by \cite{2021MNRAS.501.2790B}, utilized X-ray spectroscopy, variability, and optical/infrared counterparts. Moreover, the Galactic Bulge is a crowded region, complicating the possibility of optical spectroscopy to classify the nature of X-ray sources located there. Despite their sophistication, these surveys cover only $12\mbox{--}16\,\mathrm{deg}^2$—a negligible fraction of the sky even when combined with \textit{XMM-Newton}/Swift archival data. This highlights the critical need for all-sky surveys to detect faint QXBs.

\textit{ROSAT}'s all-sky survey delivered key data for identifying and characterizing X-ray binaries, enabling precise studies of their properties and evolution. However, the survey had limited sensitivity and relatively poor positional accuracy ($\sim30''$), hindering the identification of optical counterparts to Galactic sources \citep{1999A&A...349..389V, 2016A&A...588A.103B}. In contrast, the extended ROentgen Survey with an Imaging Telescope Array (eROSITA) survey is approximately 25 times more sensitive than \textit{ROSAT} in the soft X-ray band, and also provides significantly improved localization accuracy of about $5''$ \citep{Predehl2021}. The first eROSITA All-Sky Survey (eRASS1), publicly released on 2024 January 31, contains about one million X-ray point sources \citep{2024A&A...682A..34M}. This enables the systematic search for QXBs across the entire western Milky Way, covered by this data release. The eastern Galactic hemisphere, owned by the Russian consortium, is not yet public.

Unlike X-ray outbursts, the luminosities of QXBs are comparable to CVs and accreting white dwarfs \citep{1998ASPC..137..190K, 2018Natur.556...70H}. The population of white dwarf systems overwhelmingly outnumbers neutron star and black hole systems, making them statistically dominant in any blind search for quiescent X-ray binaries \citep{2007A&A...469..807L, 2020MNRAS.494.3799P}. The primary focus of this work is the identification of QXBs in eROSITA data, while white dwarf binaries are expected to dominate the sample due to their higher abundance. For all detected compact object binaries, the nature of the accretor can be classified through X-ray spectral analysis combined with multi-band photometric properties \citep{1998MNRAS.301..767S}.

In eROSITA DR1's X-ray sources with reliable flux ($F_\mathrm{X} > 10^{-13}\,\mathrm{erg\,cm^{-2}\,s^{-1}}$, 0.2--10\,$\mathrm{keV}$; $\sim$50 counts,  \citeauthor{2022A&A...661A..35S}\citeyear{2022A&A...661A..35S}), extragalactic sources dominate. The Galactic QXB population in this flux regime suffers significant contamination by active stars/binaries, which can be efficiently filtered using the X-ray-to-optical flux ratio ($F_\mathrm{X}/F_\mathrm{opt}$) and color criteria established by \citet{Rodriguez2024}. This methodology has proven highly effective, with \citet{2025PASP..137a4201R} identifying 23 CVs within 150 pc and a total of 590 candidate CVs within 1 kpc through precise eROSITA/Gaia DR3 astrometric matching. \citet{Wang2024} identified 444 candidate CVs and detected possible orbital modulation in 56 sources through photometric analysis of ZTF/TESS light curves. These works rely on Gaia astrometry for candidate selection, which naturally favors nearby and optically bright systems. As a result, they provide limited coverage of accreting compact object binaries. Effective photometry-only selection methods for QXBs are still under development.

In this work, we present a multi-wavelength framework to identify accreting compact object binaries in eROSITA DR1 through combined Gaia DR3 photometry and ZTF short-period light-curve analysis. Section~\ref{sec:selection} details the selection methodology, Section~\ref{sec:result} presents the results, and Section~\ref{sec:conclusion} summarizes and discusses our results.

\section{Candidates Selection} \label{sec:selection}
Our goal is to identify reliable accreting compact object binary candidates in eRASS1. To this end, we require that sources in our sample satisfy the following essential criteria:
\begin{enumerate} 
\item Accurate positional matching to an optical counterpart. 
\item X-ray emission consistent with accretion processes. 
\item Evidence of short-period photometric variability. 
\end{enumerate}
These conditions ensure the selection of candidate characteristic of accreting compact object binaries while minimizing contamination from other X-ray source populations.

\subsection{Counterparts}
\label{ssec:cp}
Optical counterpart identification in the eROSITA Final Equatorial-Depth Survey suggests that approximately 10\% of X-ray sources are Galactic \citep{2022A&A...661A...2L, 2022A&A...661A...3S}. While the typical positional uncertainty of eROSITA is $\sim$5$''$, it remains significantly larger than that of optical catalogs. As a result, the true optical counterpart is not necessarily the closest source when multiple candidates are present near an X-ray detection. However, no systematic study has yet matched optical counterparts to sources in eRASS1.

To prioritize the accuracy of counterpart identification, we adopt the following criterion: an eRASS1 source must have exactly one optical source within a 10$''$ radius.
The 10$''$ matching radius is chosen based on eROSITA's positional accuracy and statistical matching requirements. The average positional uncertainty of eRASS1 sources is typically less than 10$''$, and even smaller at high confidence levels \citep{2024A&A...682A..34M}. This radius effectively balances the need to minimize false matches while ensuring the inclusion of true optical counterparts.

The depth of an optical survey determines its ability to detect faint sources, which is critical for identifying the true optical counterparts of X-ray sources. To match the sensitivity of eROSITA, the optical catalog must have a detection limit sufficient to resolve these counterparts. Typical QXB luminosities range from \(10^{31}\) to \(10^{33}\) erg s\(^{-1}\). Using an X-ray sensitivity of \(5 \times 10^{-14}\) erg s\(^{-1}\) cm\(^{-2}\) and a luminosity of \(10^{32}\) erg s\(^{-1}\), we estimate a maximum detection distance of approximately 4 kpc. However, after correcting for interstellar absorption, the effective detection distance is expected to be smaller. At this distance, a companion star with a luminosity between 0.1 and 1 \(L_\odot\) would have an apparent magnitude of approximately 19 to 20. Consequently, we adopt Gaia DR3 as the optical catalog for counterpart identification.

\begin{figure*}[t]
    \centering
    \hfill
    \subfigure{
    \includegraphics[width=0.25\textwidth]{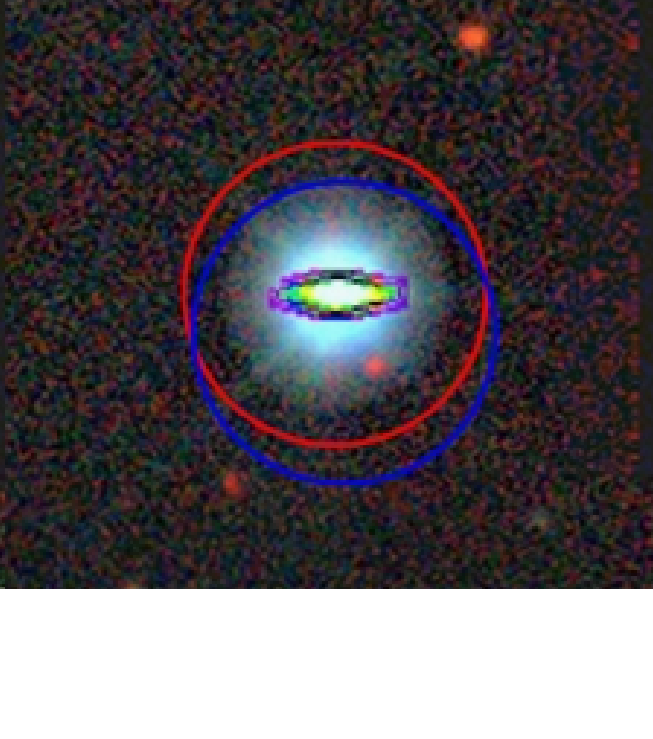}
    \label{fig:skyview}}
    \subfigure{
    \includegraphics[width=0.65\textwidth]{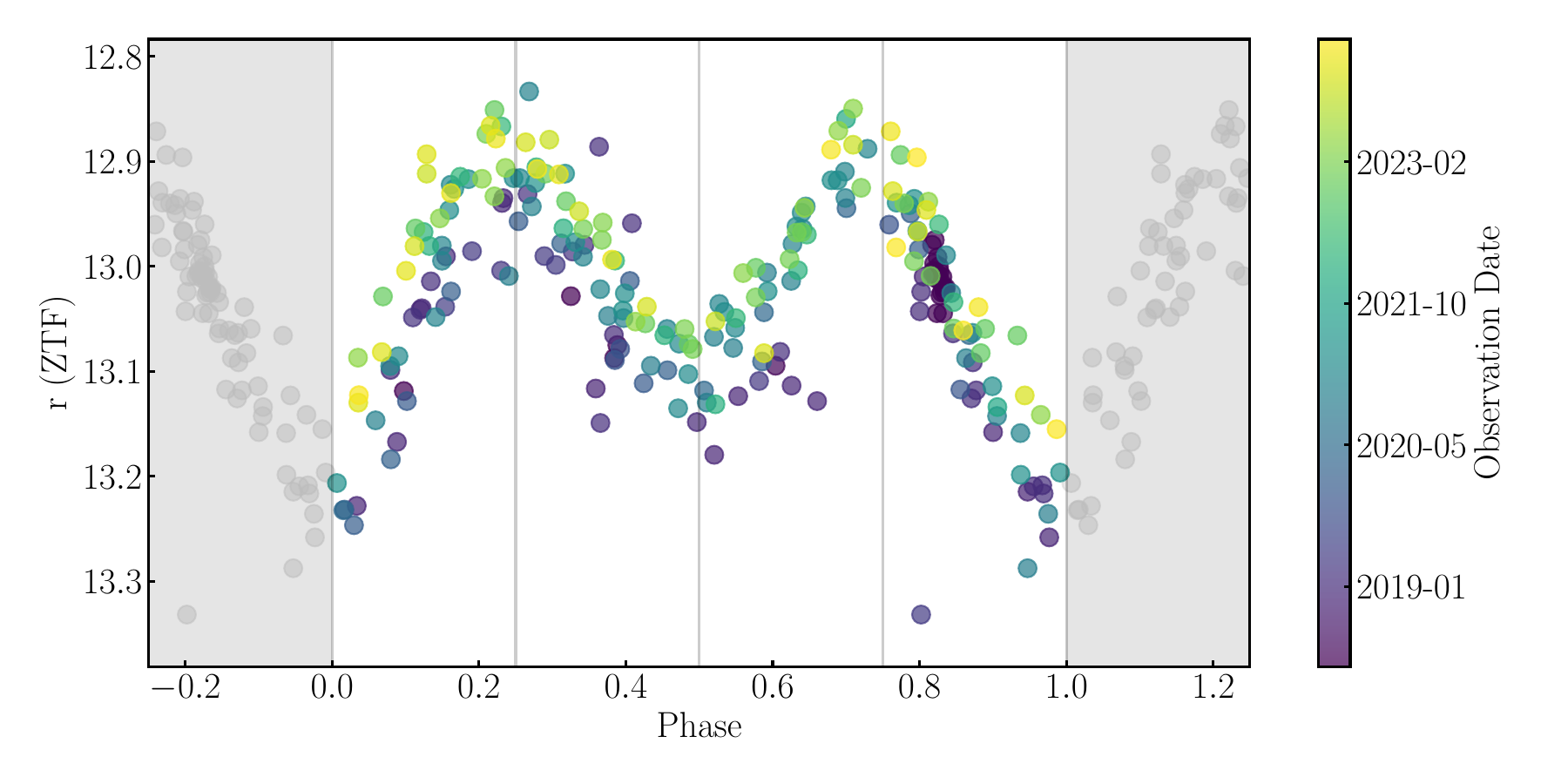}
    \label{fig:ZTFlk}}
    \subfigure{
    \includegraphics[width=0.43\textwidth]{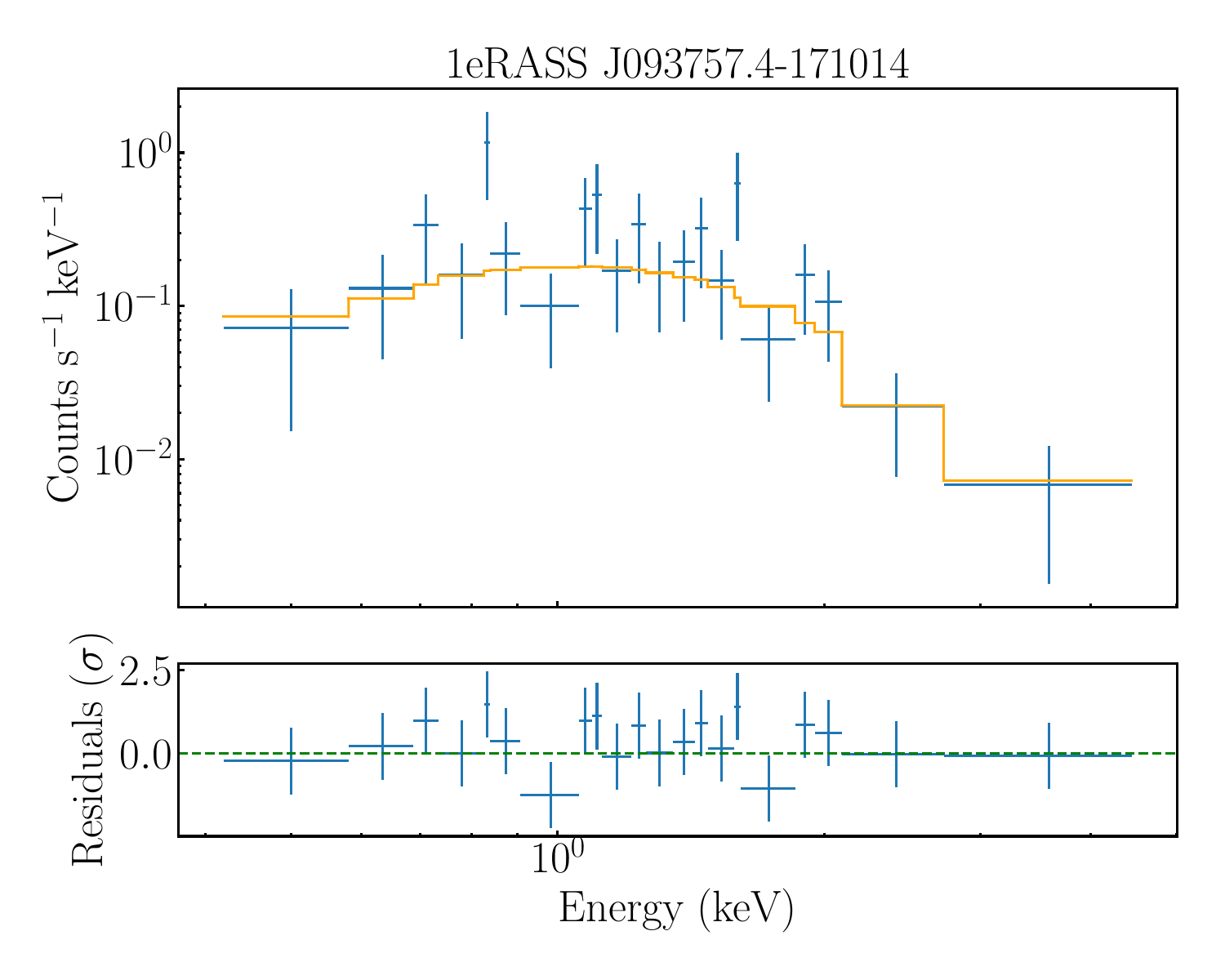}
    \label{fig:Xrayspec}}
    \hfill
    \subfigure{
    \includegraphics[width=0.54\textwidth]{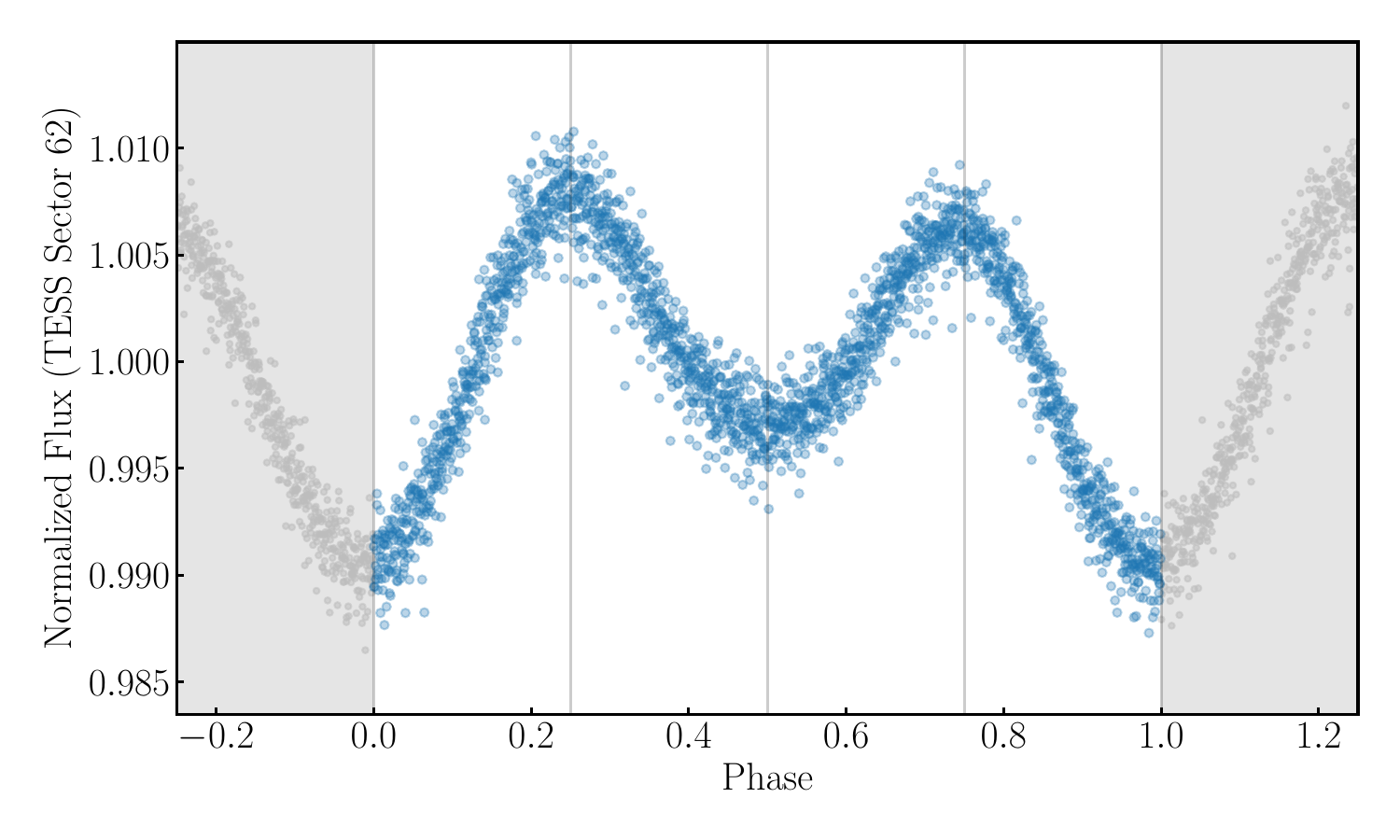}
    \label{fig:tesslk}}

    \caption{Multi-panel illustration of the candidate J093757$-$171012, shown as a representative example of our selection procedure. 
    The top-left panel (a) shows the DESI sky map with the X-ray position (blue circle) and optical counterpart (red circle) within a 10$''$ radius, with no nearby bright sources (Section~\ref{ssec:cp}). 
    The top-right panel (b) displays the ZTF phase-folded light curve, revealing ellipsoidal modulation and long-term variation (Section~\ref{ssec:lc}). 
    The bottom-left panel (c) presents the eROSITA X-ray spectrum fitted with an absorbed power-law model (\texttt{tbabs*pow}) over the 0.3--5 keV range, with a spectral index of $1.64^{+0.98}_{-0.74}$ (Section~\ref{ssec:erosita}).
    The bottom-right panel (d) shows the high-cadence TESS light curve from Sector 62, phase-folded using the ZTF period of 0.48157 days, showing clear ellipsoidal modulation (Section~\ref{ssec:lc}).
    }
    \label{fig:J0937}
\end{figure*}

Gaia DR3 provides BP/RP photometry for over 1.5 billion sources across the full sky \citep{GaiaDR3}. We cross-match eRASS1 sources with Gaia DR3 within a 10$''$ radius, identifying 367,953 X-ray sources with at least one Gaia match. We then retain only those sources with exactly one Gaia counterpart within this radius. As illustrated in Figure~\ref{fig:skyview}, a sky map of a representative source demonstrates the absence of any additional bright Gaia objects within the 10$''$ search region (blue circle) centered on the eRASS1 position.

This process yields a list of 367,953 sources matched to optical counterparts. From this, we select 127,083 targets with $\texttt{phot\_g\_mean\_mag} < 19$ for subsequent analysis.

\subsection{Distinguishing Candidates from Active Stars}
\label{gaia}

\begin{figure*}[t] 
\centering 
\includegraphics[width=0.9\textwidth]{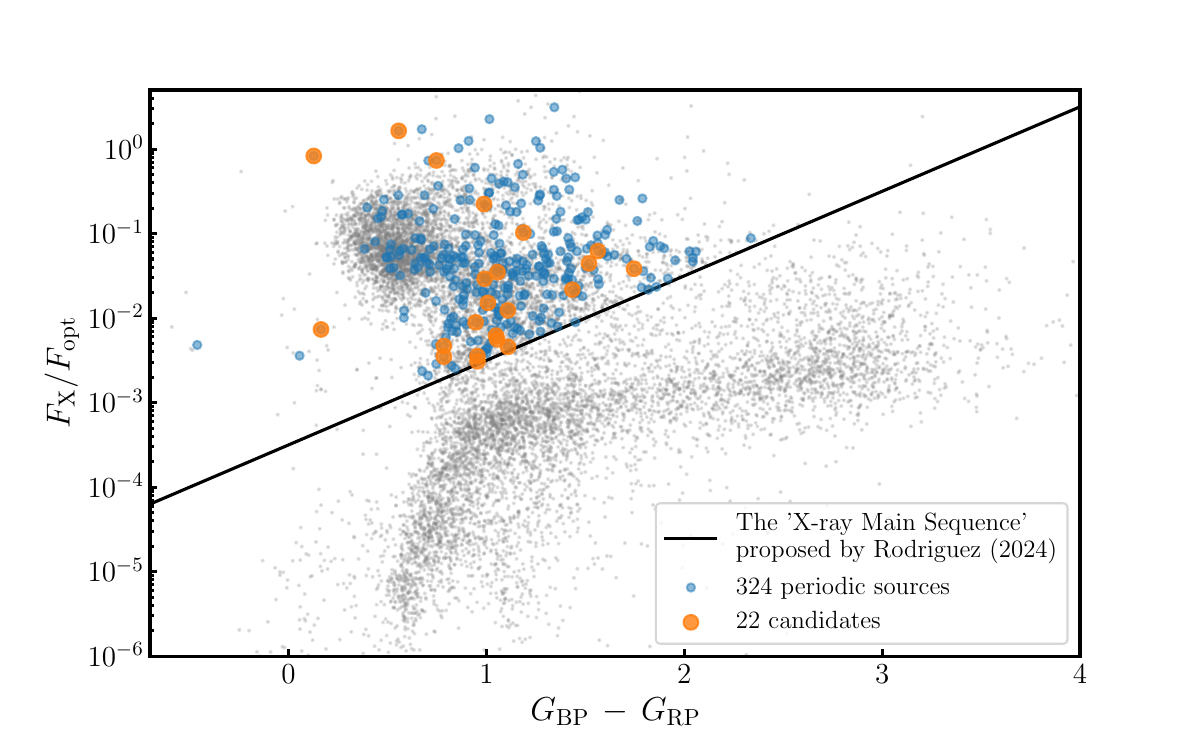} 
\caption{Distribution of the X-ray-to-optical flux ratio ($F_{\text{X}}/F_{\text{opt}}$) versus Gaia $G_{\text{BP}} - G_{\text{RP}}$ color for eRASS1 sources matched with Gaia DR3 counterparts. The gray background shows all 127,083 X-ray sources, blue points represent 324 potential candidates identified based on flux ratios and peak Lomb-Scargle powers, and orange points highlight 22 candidates exhibiting clear short-period light curves. The black line indicates the empirical cut at $\log(F_{\text{X}}/F_{\text{opt}}) = (G_{\text{BP}} - G_{\text{RP}}) - 3.5$, used to distinguish accreting systems from magnetically active stars.} \label{fig:el} 
\end{figure*}

Magnetically active stars constitute a dominant class of Galactic X-ray sources \citep{2002ApJ...574..258F, 2009ApJS..181..444A, 2017ARA&A..55..261K}. To distinguish X-ray emission due to stellar activity from that produced by accretion, we use Gaia photometry in conjunction with the X-ray-to-optical flux ratio. For the 127,083 sources with Gaia counterparts, we retrieve $G$, $G_{\text{BP}}$ and $G_{\text{RP}}$ from Gaia DR3 (\texttt{phot\_g\_mean\_mag}, \texttt{phot\_bp\_mean\_mag} and \texttt{phot\_rp\_mean\_mag}), along with X-ray fluxes in the 0.2--2.3 keV band from eROSITA DR1 (\texttt{ML\_FLUX\_1}).  These values are plotted in the $F_{\text{X}}/F_{\text{opt}}$ vs. $G_{\text{BP}} - G_{\text{RP}}$ diagram shown in Figure~\ref{fig:el}.

To identify accretion-powered systems, we adopt the empirical limit proposed by \citet{Rodriguez2024}, which is based on \textit{XMM-Newton} 0.2-12 keV band data: $\log(F_\mathrm{X}/F_\mathrm{opt}) > G_{\text{BP}} - G_{\text{RP}} - 3.5$. Here, the optical flux is calculated as $F_{\mathrm{opt}} = 10^{-0.4\,(G - 4.83)}\,L_{\odot}\,/(4\pi \,(10\,\mathrm{pc})^{2})$, and the X-ray flux is used directly from (\texttt{ML\_FLUX\_1}) in 0.2-2.3 keV band. Assuming a power-law with an index of 2, our empirical boundary is approximately 1.8 times higher than that reported in \cite{Rodriguez2024}. Consequently, applying our selection criteria imposes a more stringent constraint on our sample. Applying this criterion yields 57,940 sources with flux ratios above the activity threshold.

Given the known discrepancies between Gaia DR3 astrometric solutions and dynamical constraints for some low mass X-ray binaries (LMXBs), we do not impose cuts on astrometric parameters.  As a result, the sample includes a number of extragalactic X-ray sources such as AGNs and quasars. However, these contaminants are effectively removed in the subsequent step based on short-period optical variability.

\subsection{ZTF Light Curves}
\label{ssec:lc}
Photometric light curves are a powerful diagnostic for detecting periodic variability in QXBs. In such systems, tidal distortion of the companion can induce ellipsoidal modulation, while other features, such as eclipses, reflection effects, or stochastic accretion variability, may complicate or obscure the ellipsoidal modulation. Despite these complexities, time-domain analysis remains essential for identifying periodic behavior indicative of accreting compact object binaries.

The Zwicky Transient Facility (ZTF) is a wide-field optical survey covering the northern sky to a depth of $\sim$20.5 (r band, 5$\sigma$) \citep{2019PASP..131a8002B}.  Its 22nd public data release (DR22) includes $\sim$58.5 million images and $\sim$4.92 billion light curves. The ZTF collection of light curve data is accessed through the Application Programming Interface (API) provided by Infrared Science Archive (IRSA)\footnote{\url{https://irsa.ipac.caltech.edu/docs/program_interface/ztf_lightcurve_api.html}}. For each of the 57,940 X-ray-selected candidates, we query g- and i-band photometry within a 5$''$ radius of the Gaia position using the IRSA API. Of the 57,940 sources, 17,221 match ZTF DR22 time-domain photometry, yielding 33,048 light curves in two bands.

We employ the Lomb-Scargle algorithm to search for periodic signals \citep{1976Ap&SS..39..447L,1981ApJS...45....1S}, folding the light curves with the obtained periods $T_\mathrm{ph}$. The period search window is set to 0.1--4 days, with frequency boundaries defined as \texttt{minimum\_frequency} = 0.25 and \texttt{maximum\_frequency} = 10. This choice is motivated by our focus on X-ray binaries, which generally have longer periods than two hours. For each candidate, we fold the light curves using the period of maximum power and visually inspect both the power spectrum and phase-folded light curve. This yields 324 candidates exhibiting significant periodic modulation.

These 324 candidates are then visually inspected to eliminate those with periods likely induced by Earth's rotation (e.g., period very close to 1 day) or improper folding. Ultimately, we select 22 targets. While their light curves may exhibit contamination or additional components, the dominant variability in each case corresponds to the period of the peak power in the Lomb-Scargle spectrum. An example of a periodic light curve is shown in Figure~\ref{fig:ZTFlk}.

\section{Results}
\label{sec:result}
We identify 22 accreting compact object binary candidates in eROSITA DR1 through a three-step selection process: (1) cross-matching to a unique Gaia optical counterpart within a 10$''$ radius; (2) requiring X-ray-to-optical flux ratios exceeding the empirical threshold for stellar activity; and (3) detecting short-period variability in ZTF time-domain photometry. Among these candidates, two are previously known systems: J041920$+$072545 (hereafter J0419) and J102347$+$003840 (hereafter J1023). J0419 was confirmed via dynamical analysis using data from the Large Sky Area Multi-Object Fiber Spectroscopic Telescope (LAMOST, \citeauthor{2012RAA....12.1197C} \citeyear{2012RAA....12.1197C}) as a binary comprising an extremely low-mass pre-white dwarf and a compact object with a mass of $1.09 \pm 0.05M_{\odot}$ \citep{2022ApJ...933..193Z}. J1023 is a transitional millisecond pulsar, exhibiting a bistable accretion behavior, switching between a radio pulsar state and a LMXB phase depending on the mass transfer rate \citep{2009Sci...324.1411A}. A summary of all 22 candidates is provided in Table~\ref{tab:1}.

\subsection{Archival Spectroscopic Survey Data Matching}

\begin{figure*}[t] 
\centering \includegraphics[width=0.9\textwidth]{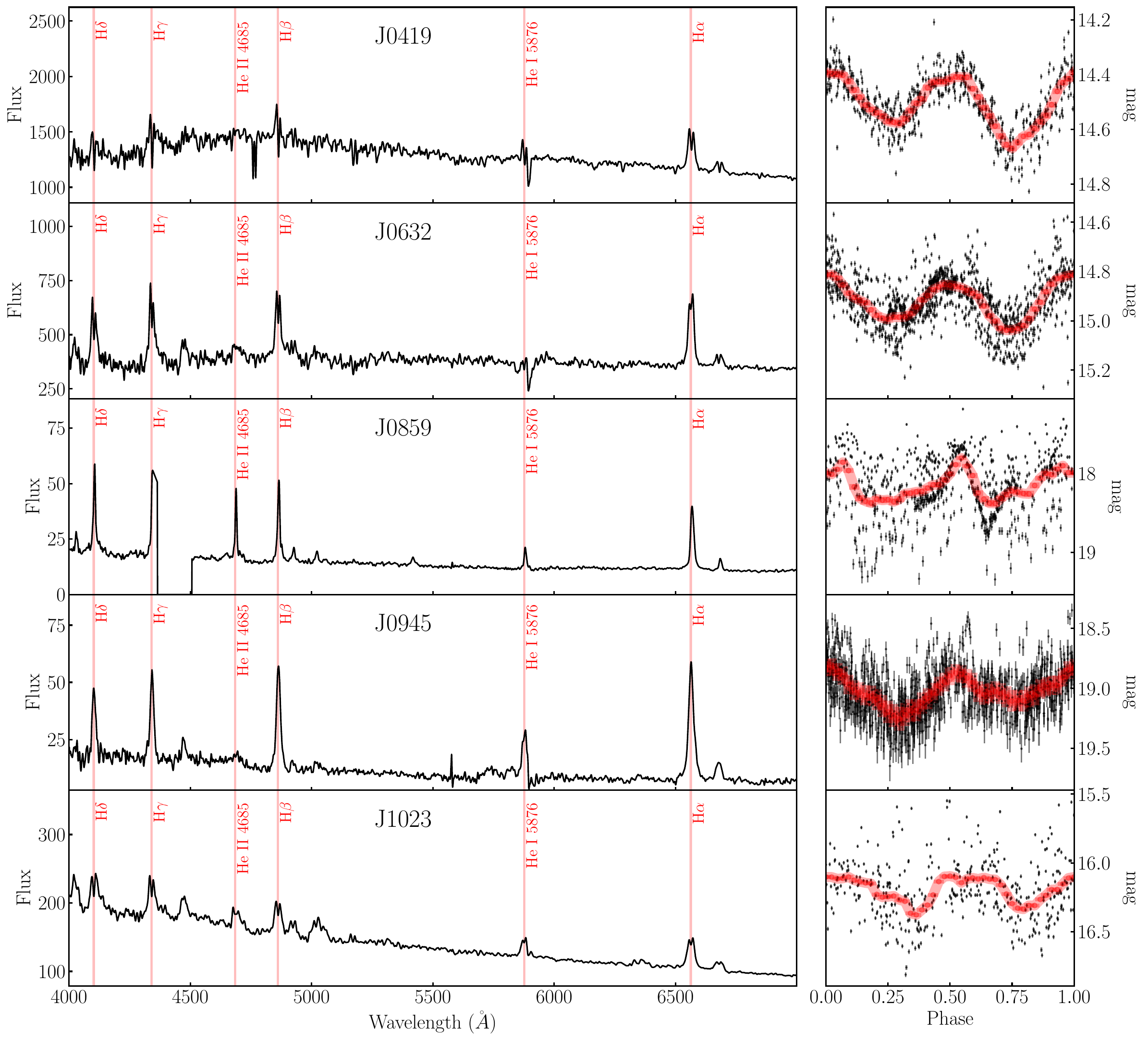} 
\caption{Spectroscopic and photometric properties of five representative candidates (J0419, J0632, J0859, J0945, and J1023) with available optical spectra from the SDSS or LAMOST databases.
    Left panels: Optical spectra from SDSS or LAMOST, showing prominent Balmer and He\,\textsc{I}/He\,\textsc{II} emission lines, consistent with accreting compact object binaries.
    Right panels: Phase-folded ZTF $r$-band light curves. The phase-folded light curves exhibit orbital modulation, although the variability amplitudes appear to be diluted, possibly due to contamination from the accretion disk.} \label{fig:spectra}
\end{figure*}

We cross-match our candidates with the LAMOST and Sloan Digital Sky Survey (SDSS, \citeauthor{2020ApJS..249....3A} \citeyear{2020ApJS..249....3A}). LAMOST provides time-domain spectroscopy of J0419, and reveals double-peaked emission lines indicative of an accretion disk in spectra. Additionally, LAMOST conducted low-resolution spectroscopic observations of J063213$+$253622 (hereafter J0632), J085909$+$053654 (hereafter J0859), and J094558$+$292252 (hereafter J0945), which are classified as CVs in the LAMOST Low Resolution Survey. The SDSS database includes spectroscopic observations of J1023, J0859, and J0945. As shown in Figure~\ref{fig:spectra}, the spectrum of J1023 exhibits double-peaked emission lines from an accretion disk, while J0859 and J0945 show emission line features characteristic of CVs. Figure~\ref{fig:gaia} displays these five sources as four blue circles and one red circle, which generally correspond to the uppermost sources in the diagram, thereby indicating that they are located furthest from the X-ray main sequence towards the upper-left region in Figure~\ref{fig:el}. No spectroscopic data from either LAMOST or SDSS are available for the remaining 17 candidates. The catalog of candidate CVs out to 1000 pc by \cite{2025PASP..137a4201R} includes four systems in our sample: J0419, J0632, J0859, and J0945. The remaining systems are not included.

\subsection{SIMBAD Classifications of Candidates}

\begin{figure*}[t] 
\centering \includegraphics[width=0.9\textwidth]{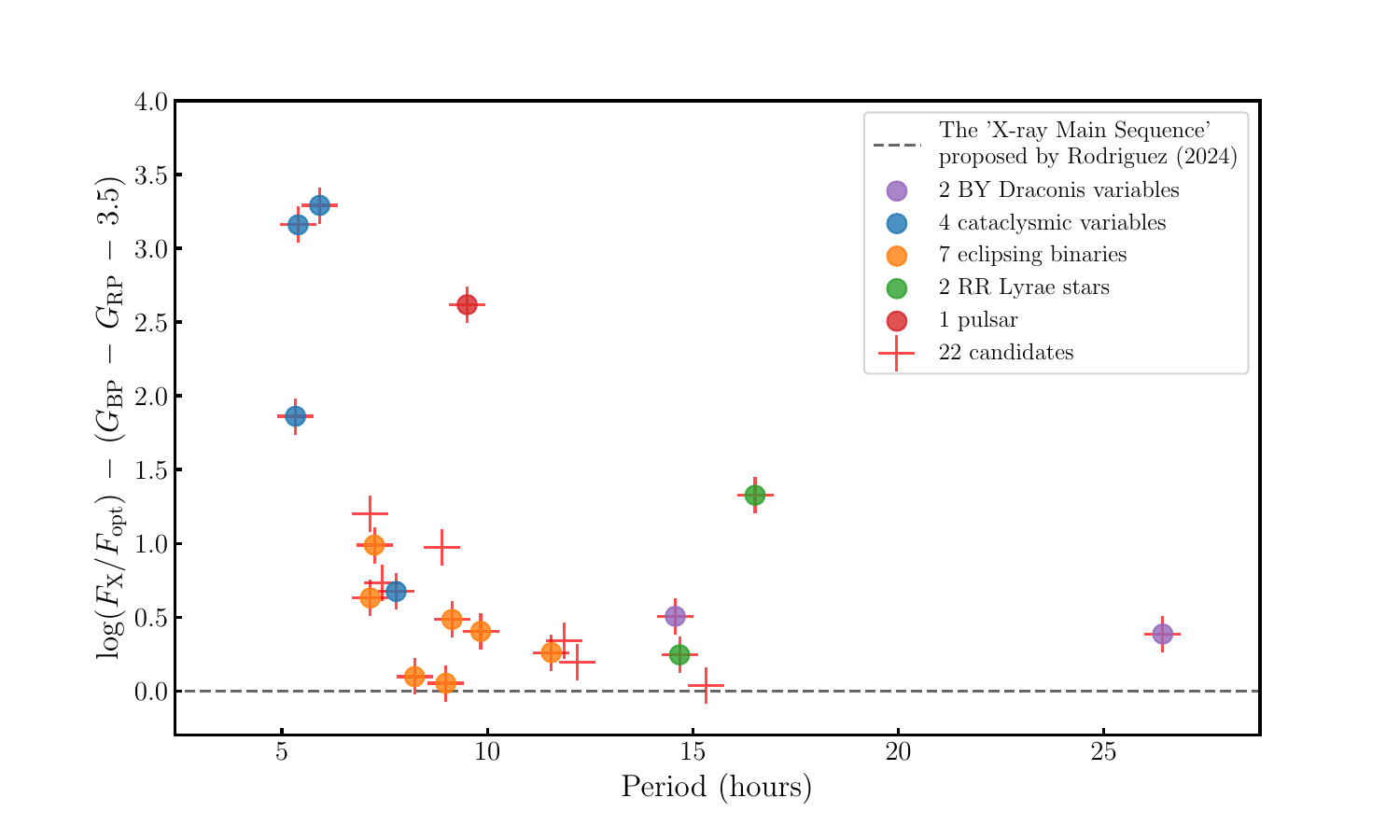} 
\caption{Offset from the empirical cut in Figure~\ref{fig:el} as a function of orbital period for all 22 candidates. 
The $y$-axis shows the vertical distance from the empirical relation $\log(F_X/F_{\mathrm{opt}}) = (G_{\text{BP}} - G_{\text{RP}}) - 3.5$ (the dashed line), 
while the $x$-axis corresponds to the orbital period. 
All 22 candidates are represented by red crosses. 
Colored circles indicate sources with classifications retrieved from SIMBAD, where the light curve types are based on references from Gaia Early Data Release 3 (EDR3) or ZTF. 
These classifications were not assigned in this work.
} \label{fig:gaia}
\end{figure*}

We searched for SIMBAD entries corresponding to our 22 candidates by querying their optical coordinates. 
SIMBAD has adopted classifications for 15 of 22  candidates, including the well-studied accreting compact object binaries J0419 and J1023. 
Among them, 12 classifications are based on sources from Gaia EDR3, as illustrated in Figure~\ref{fig:gaia}. 
According to Gaia EDR3, J0632, J081210$+$040351, J0859, and J0945 are classified as CVs; 
J051723$-$115359 and J052316$-$252737 as RR Lyrae stars; and 
J064513$+$195511, J073223$-$072292, J073644$+$131707, J080038$-$275245, J093757$-$171012 (hereafter J0937), and J094146$+$060933 as eclipsing binaries. 
Additionally, ZTF classified J0419 and J152611$-$102512 as BY Draconis variables. 
The remaining 7 candidates currently have no classification or identification recorded in SIMBAD.

\subsection{eROSITA X-ray Analysis}
\label{ssec:erosita}
For candidates with X-ray fluxes exceeding the threshold $\texttt{ML\_FLUX\_1} > 10^{-13}$ $\mathrm{erg\,cm^{-2}\,s^{-1}}$, we retrieve their source detection products to fit the X-ray spectra. Using the \texttt{grppha} tool, we group the spectra with a minimum of three channels per bin. We fit the spectra in the 0.3--5 keV range using a simple absorbed power-law model, \texttt{tbabs*pow}, and estimate the unabsorbed flux in the 0.5--10 keV band using the \texttt{cflux} command in XSPEC (version 12.12.0). The \texttt{tbabs} model \citep{2000ApJ...542..914W} employs photoelectric cross-sections from \citet{1996ApJ...465..487V}. All reported errors are at the 1$\sigma$ confidence level.
An example of the fitting on X-ray spectrum is shown in Figure~\ref{fig:Xrayspec}.

\begin{figure*}[t]
    \centering
    \subfigure{
        \includegraphics[width=0.48\textwidth]{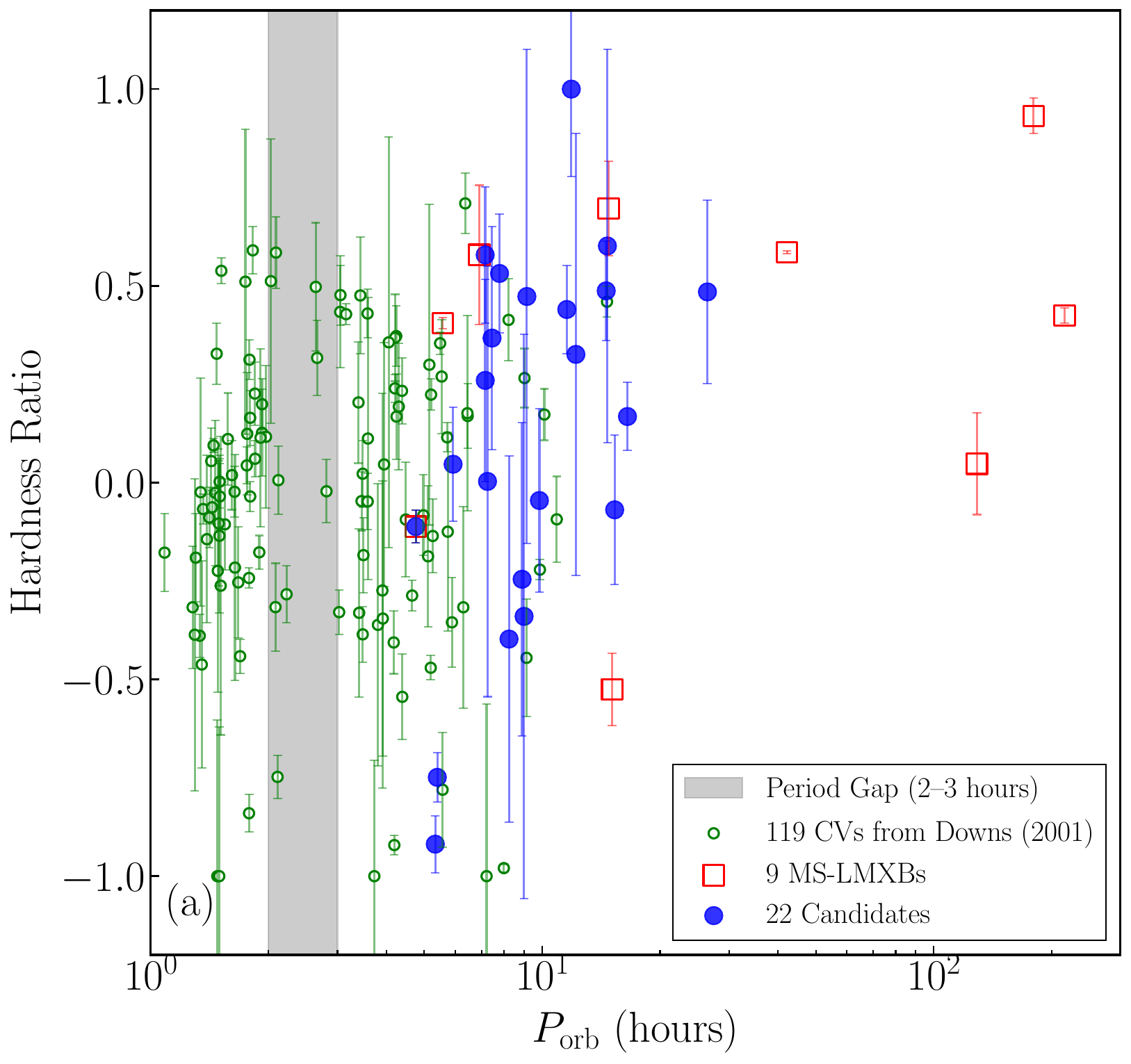}
        \label{fig:hr}}
    \hfill
    \subfigure{
        \includegraphics[width=0.47\textwidth]{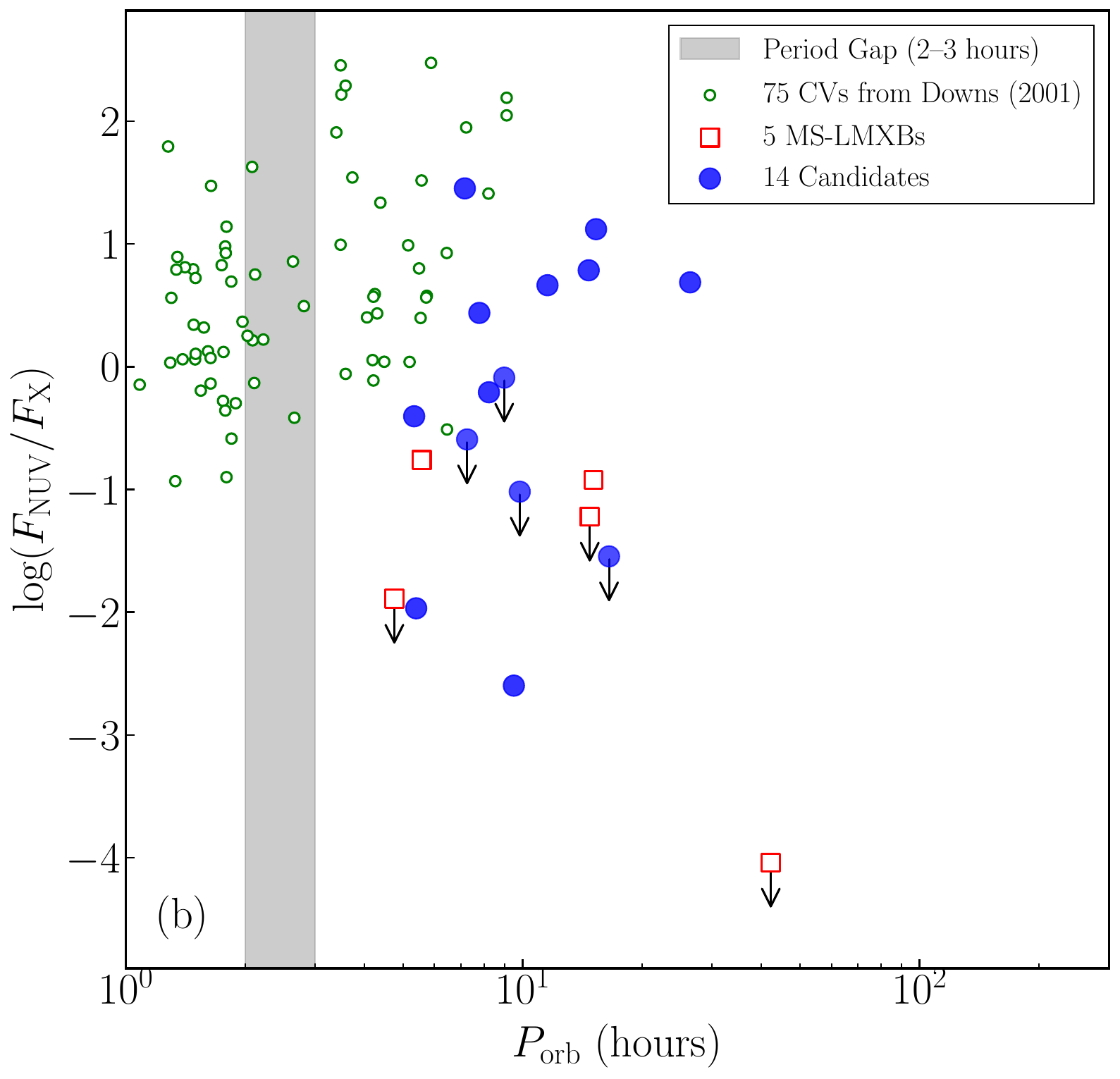}
        \label{fig:uv}}
    \caption{
    Multi-wavelength properties of the accreting compact object binary candidates, plotted against orbital period. Panel (a) displays the hardness ratio, and panel (b) shows the NUV-to-X-ray flux ratio ($F_{\mathrm{NUV}}/F_{\mathrm{X}}$) derived from the \textit{GALEX} NUV magnitude and the \texttt{ML\_FLUX\_1} measurement in eRASS1. 
    In both panels, marker styles are used consistently: blue circles represent our candidates, green circles denote CVs, and red squares mark known MS-LMXBs. The legends summarize these classifications visually.
    Arrows in the right panel denote $5\sigma$ NUV flux upper limits where no significant NUV flux was detected. 
    The $x$-axis in both panels corresponds to the orbital period in hours. 
    }
    \label{fig:4}
\end{figure*}

The limited photon counts hinder robust spectral analysis, resulting in large uncertainties in the fitted parameters and reduced reliability of the \texttt{tbabs*pow} model. This scarcity of data prevents detailed characterization of the X-ray emission properties for most candidates. Future observations with longer exposure times or joint analyses with other X-ray instruments could mitigate these limitations, enabling more precise spectral modeling and deeper insights into the nature of these systems.

In the absence of sufficient photon counts for full spectral fitting of 16 candidates, we use fluxes from the eROSITA DR1 catalog to construct hardness ratios. Source count rates are retrieved in four energy bands, with \texttt{ML\_RATE\_P1}, \texttt{ML\_RATE\_P2}, \texttt{ML\_RATE\_P3}, and \texttt{ML\_RATE\_P4} respectively representing the 0.2 to 0.5 keV, 0.5 to 1 keV, 1 to 2 keV, and 2 to 5 keV ranges. In subsequent calculations, we denote these rates as P1, P2, P3, and P4. We calculate a hardness ratio (HR) for each source to characterize its spectral properties. The hardness ratio is defined as:
 \[\mathrm{HR} = \frac{(P4 + P3) - (P2 + P1)}{P4 + P3 + P2 + P1}\]
We then give the distribution of HR as a function of orbital period in Figure~\ref{fig:hr}.

To compare our candidates with established CV and LMXB samples, we performed two additional cross-matching analyses. First, we cross-matched the eRASS1 with the catalog of 339 known LMXBs compiled by \citet{2024A&A...684A.124F}, and selected systems that (1) have measured orbital periods and (2) have companion stars classified as stellar spectral types (i.e., non-degenerate stars). Specifically, we excluded systems with white dwarf companions and retained only those with main-sequence-like donors. This yielded a reference set of 9 LMXBs with main-sequence companions (hereafter MS-LMXBs for convenience). Second, we cross-matched the eRASS1 with the CV catalog from \citet{2001PASP..113..764D} and selected those with available orbital period measurements, resulting in 119 CVs. As shown in Figure~\ref{fig:hr}, our candidates (blue circles) are plotted alongside comparison samples of MS-LMXBs (red squares) and CVs (green open circles).

Compared to CVs, which usually have orbital periods below 10 hours, LMXBs can extend to significantly longer periods. \cite{1998A&AS..132..341M} ’s analysis of X-ray sources in the ROSAT survey indicates that CVs exhibit systematically softer X-ray emission than neutron star or black hole systems. CVs are expected to occupy the lower-left region of Figure~\ref{fig:hr}, characterized by short orbital periods and relatively soft X-ray emissions. In contrast, neutron star or black hole binaries generally exhibit harder X-ray emission. The distribution of our 22 candidates is broadly consistent with this trend, with the lower envelope showing a positive correlation between HR and orbital period. This suggests that the combination of X-ray hardness and orbital period may provide a useful empirical tool for distinguishing CVs from more compact accretors such as neutron stars and black holes.

\subsection{GALEX Ultraviolet Data}
\label{ssec:uv}
We utilize the Galaxy Evolution Explorer (GALEX) Data Release 7 (DR7) to match ultraviolet photometry for our 22 candidates. Through cross-matching the GALEX DR7 catalog with our sample within a 5$''$ radius, we obtain NUV magnitudes for seven sources, including J0632 and J0859. Additionally, we query the Multi-mission Archive at STScI (MAST) for GALEX tiles upon our candidates and download the associated Coadded Imaging Products. The Merged Catalog of Sources (\texttt{AIS\_$\ast$\_mcat.fits}) provides NUV magnitudes for other six sources. We note that some faint sources (NUV $\sim$ 23 in the merged catalog) have fluxes approaching the background noise level. For four faint candidates, we derive flux upper limits of 5$\sigma$ using the original Imaging files (\texttt{AIS\_$\ast$\_int.fits}). Figure~\ref{fig:uv} illustrates the distribution of the NUV-to-X-ray flux ratio versus orbital period for our candidates. The observed flux ratios span several orders of magnitude. We have similarly cross-matched the CVs and MS-LMXBs appearing in Figure~\ref{fig:hr} with the GALEX DR7 catalog, and plotted them on the figure using the same markers for direct comparison. As with X-ray hardness ratios, differences in accretion physics between neutron star/black hole and white dwarf systems may also appear in softer-band flux ratios, such as NUV-to-X-ray.
In principle, this ratio could serve as a boundary between the two populations. However, due to the limited depth of GALEX and the limited size of known sample, a comprehensive analysis has yet to be carried out. With more classified white dwarf and neutron star/black hole systems, we anticipate a clearer separation in this parameter space. This will facilitate more effective classification of candidates in our future work.

\begin{deluxetable}{cccccccccc}
\tablecaption{Full Candidates\label{tab:1}}
\tablehead{
\colhead{Gaia Source} &
\colhead{$\mathrm{Plx}$} &
\colhead{$\mathrm{RPlx}$} &
\colhead{$G$} &
\colhead{$G_{\text{BP}} - G_{\text{RP}}$} &
\colhead{$F_{\mathrm{X}}$} &
\colhead{$L_{\mathrm{X}}$} &
\colhead{$T_\mathrm{ph}$} &
\colhead{$\Gamma$} \\
\colhead{} &
\colhead{(mas)} &
\colhead{} &
\colhead{} &
\colhead{} &
\colhead{($\mathrm{erg\,cm^{-2}\,s^{-1}}$)} &
\colhead{($\mathrm{erg\,s^{-1}}$)} &
\colhead{(day)} &
\colhead{} &
}
\startdata
J041920$+$072545 & 1.45 & 52.38 & 14.7 & 0.95 & $3.36 \times 10^{-13}$ & $1.93 \times 10^{31}$ & 0.607 & $2.44_{-1.22}^{+1.61}$ \\
J051723$-$115359 & 0.48 & 7.88 & 17.01 & 1.05 & $2.76 \times 10^{-14}$ & $1.45 \times 10^{31}$ & 0.612 & --  \\
J052316$-$252737 & 0.45 & 10.87 & 16.53 & 1.19 & $7.18 \times 10^{-13}$ & $4.33 \times 10^{32}$ & 0.688 & $1.45_{-0.20}^{+0.37}$ \\
J060502$-$243748 & 1.70 & 124.21 & 13.84 & 0.96 & $2.54 \times 10^{-13}$ & $1.06 \times 10^{31}$ & 0.638 & -- \\
J063213$+$253622 & 2.24 & 73.67 & 15.11 & 1.01 & $3.91 \times 10^{-13}$ & $9.39 \times 10^{30}$ & 0.324 & -- \\
J064513$+$195511 & 1.49 & 23.26 & 16.52 & 1.52 & $3.12 \times 10^{-13}$ & $1.69 \times 10^{31}$ & 0.298 & -- \\
J071307$-$172149 & 0.45 & 4.04 & 17.98 & 1.75 & $7.02 \times 10^{-14}$ & $4.15 \times 10^{31}$ & 0.495 & -- \\
J072518$+$081255 & 0.77 & 15.91 & 16.22 & 1.05 & $5.17 \times 10^{-14}$ & $1.05 \times 10^{31}$ & 0.508 & -- \\
J073223$-$072292 & 0.33 & 6.21 & 16.45 & 1.11 & $3.41 \times 10^{-14}$ & $3.81 \times 10^{31}$ & 0.375 & -- \\
J073644$+$131707 & 0.61 & 20.82 & 15.35 & 0.95 & $7.33 \times 10^{-14}$ & $2.38 \times 10^{31}$ & 0.343 & -- \\
J074236$+$020118 & 0.63 & 23.50 & 15.06 & 0.16 & $1.98 \times 10^{-13}$ & $5.95 \times 10^{31}$ & 0.298 & -- \\
J074332$+$061045 & 0.92 & 6.49 & 17.80 & 1.56 & $1.34 \times 10^{-13}$ & $1.89 \times 10^{31}$ & 0.310 & -- \\
J080038$-$275245 & 0.34 & 4.16 & 17.61 & 1.11 & $3.18 \times 10^{-14}$ & $3.26 \times 10^{31}$ & 0.381 & -- \\
J081210$+$040351 & 0.54 & 2.83 & 18.69 & 0.56 & $1.56 \times 10^{-12}$ & $6.40 \times 10^{32}$ & 0.225 & $3.53_{-0.39}^{+0.72}$ \\
J085909$+$053654 & 2.25 & 19.01 & 18.07 & 0.99 & $3.78 \times 10^{-13}$ & $8.95 \times 10^{30}$ & 0.222 & -- \\
J093757$-$171012 & 2.48 & 132.87 & 13.09 & 0.79 & $5.77 \times 10^{-13}$ & $1.13 \times 10^{31}$ & 0.482 & $1.64_{-0.74}^{+0.98}$ \\
J094146$+$060933 & 1.54 & 27.81 & 15.98 & 1.43 & $2.49 \times 10^{-13}$ & $1.26 \times 10^{31}$ & 0.410 & -- \\
J094558$+$292252 & 2.65 & 9.44 & 18.98 & 0.13 & $6.03 \times 10^{-13}$ & $1.03 \times 10^{31}$ & 0.247 & $1.15_{-0.46}^{+1.51}$ \\
J102347$+$003840 & 0.69 & 9.72 & 16.23 & 0.75 & $6.70 \times 10^{-12}$ & $1.71 \times 10^{33}$ & 0.396 & $2.24_{-0.23}^{+0.26}$ \\
J134752$-$263155 & 0.45 & 2.61 & 18.18 & 0.99 & $4.45 \times 10^{-14}$ & $2.63 \times 10^{31}$ & 0.370 & -- \\
J152611$-$102512 & 0.59 & 16.93 & 15.16 & 0.79 & $1.14 \times 10^{-13}$ & $3.91 \times 10^{31}$ & 1.101 & -- \\
J073523$+$123009 & -0.01 & -0.06 & 17.90 & 1.06 & $6.86 \times 10^{-14}$ & $1.46 \times 10^{35}$ & 0.302 & --  \\
\enddata
\tablecomments{
This table summarizes the properties of 22 accreting  compact object binary candidates. Columns include: Gaia source designation; parallax ($\mathrm{Plx}$) and parallax over error ($\mathrm{RPlx}$); Gaia G-band magnitude ($G$) and color ($G_{\text{BP}} - G_{\text{RP}}$); X-ray flux ($F_{\mathrm{X}}$) from eROSITA; X-ray luminosity ($L_{\mathrm{X}}$)  estimated from $\mathrm{Plx}$ and $F_{\mathrm{X}}$; photometric period ($T_\mathrm{ph}$) from ZTF light curves; X-ray photon index ($\Gamma$) from power-law fitting to the eROSITA spectrum. A dash (--) indicates insufficient photon counts to fit.}
\end{deluxetable}

\section{Conclusions and Discussion}
\label{sec:conclusion}
Combining Gaia DR3 and ZTF time-domain photometry, we search for accreting compact object binaries in the eRASS1 catalog. We focus on those eRASS1 X-ray sources, which match only one Gaia counterparts within a 10$''$ radius. Using the X-ray-to-optical flux ratio and Gaia color, we select the X-ray sources whose X-ray emission dominately produced by accretion. We identify 324 sources exhibiting periodic light curves by searching for periodic signals in ZTF time-domain photometry. Finally, we obtain 22 accreting compact object binary candidates after visual inspection.
Among our candidates, two well-studied accreting compact object binaries are included (J0419 and J1023). J0419 is an extremely low mass pre-whtie dwarf binary system, and J1023 is a neutron star X-ray binary. 

We collect spectroscopic data from LAMOST and SDSS for five sources: J0419,  J0632, J0859, J0945 and J1023. J0419 and J1023 have multiple spectra, with features consistent with their established classifications. J0632, J0859, and J0945 exhibit spectral characteristics of CV and are classified as CVs by LAMOST and/or SDSS. The Gaia EDR3 catalog provides photometric classifications for 12 additional sources, including CVs, RR Lyrae stars, and eclipsing binaries, as shown in Figure~\ref{fig:el}. We fit eROSITA X-ray spectra using absorbed power-law model, with J0937 exhibiting a relatively hard spectrum, as shown in Figure~\ref{fig:Xrayspec}. 

Reflecting on the selection process, requiring a unique Gaia optical counterpart within 10$''$ is a stringent criterion. In the absence of an official eRASS1 optical counterpart catalog, this approach effectively identifies accurate optical counterparts, though it excludes a significant number of X-ray sources. As an initial effort in our series of studies, we plan to adopt more relaxed matching criteria in future searches. When excluding active stars, we focus on the X-ray-to-optical flux ratio rather than absolute flux values, simplifying calculations such as X-ray flux conversions. However, this simplification shifts the empirical relation in Figure~\ref{fig:el}, potentially excluding one additional source. We retain the original boundary in our results, noting that sources near this boundary warrant further analysis. Nevertheless, the adopted boundary effectively distinguishes active stars from accretion-powered systems. Overall, we adopted a relatively stringent selection strategy throughout the filtering process, which excluded the majority of sources and yielded only 22 candidates out of over a million initial targets. With the future release of time-domain observations from WFST and future eROSITA data releases, our approach is expected to become increasingly effective in the search for accreting compact object binaries.

In selecting ZTF periodic variability, we retain sources with clear light curves regardless of their variability patterns. Accretion disks are luminous and highly variable, producing complex light curves with features resembling eclipses or pulsations. However, the prior exclusion of active stars using the X-ray-to-optical flux ratio significantly reduces the likelihood of contamination by eclipsing binaries or pulsating variables. The short-period variability sources we identify are highly promising X-ray binary candidates. We obtained spectroscopic data of J0937 using the  CFHT, aiming to confirm its accreting nature; the data are currently being reduced. Spectroscopic confirmation of additional reported candidates will be carried out progressively in future observations.

We initiated this study with Gaia DR3 as the optical catalog, representing a preliminary effort to mine accreting compact object binaries from eROSITA data. However, as demonstrated throughout this work, our selection does not rely on Gaia astrometric parameters, as no parallax, proper motion, or RUWE cuts were applied. Importantly, the only requirement on the optical catalog is the availability of broad-band photometry to compute colors and flux ratios. This makes our approach fully adaptable to other other wide-field photometric surveys with enough depth and coverage\, such as the Rubin Observatory LSST. In conjunction with X-ray-to-optical flux ratio and time-domain photometry, this strategy offers a scalable pathway to uncover quiescent LMXBs and CVs beyond Gaia's astrometric reach.

\section*{acknowledgments}
We thank the anonymous referee for the constructive suggestions to improve the paper. This work was supported by the National Natural Science Foundation of China under grants 12433007, 12221003 and 12393814.
J.Z.L was supported by the Tianshan Talent Training Program through the grant 2023TSYCCX0101. We acknowledge the science research grants from the China Manned Space Project with No. CMS-CSST-2025-A13.
This work is based on data from eROSITA, the soft X-ray instrument aboard SRG, a joint Russian-German science mission supported by the Russian Space Agency (Roskosmos), in the interests of the Russian Academy of Sciences represented by its Space Research Institute (IKI), and the Deutsches Zentrum für Luft- und Raumfahrt (DLR). The SRG spacecraft was built by Lavochkin Association (NPOL) and its subcontractors, and is operated by NPOL with support from the Max Planck Institute for Extraterrestrial Physics (MPE). The development and construction of the eROSITA X-ray instrument was led by MPE, with contributions from the Dr. Karl Remeis Observatory Bamberg \& ECAP (FAU Erlangen-Nuernberg), the University of Hamburg Observatory, the Leibniz Institute for Astrophysics Potsdam (AIP), and the Institute for Astronomy and Astrophysics of the University of Tübingen, with the support of DLR and the Max Planck Society. The Argelander Institute for Astronomy of the University of Bonn and the Ludwig Maximilians Universität Munich also participated in the science preparation for eROSITA.
ZTF is a public-private partnership, with equal support from the ZTF Partnership and from the U.S. National Science Foundation through the Mid-Scale Innovations Program (MSIP).

\bibliography{main}{}

\begin{thebibliography}{}
\expandafter\ifx\csname natexlab\endcsname\relax\def\natexlab#1{#1}\fi
\providecommand{\url}[1]{\href{#1}{#1}}
\providecommand{\dodoi}[1]{doi:~\href{http://doi.org/#1}{\nolinkurl{#1}}}
\providecommand{\doeprint}[1]{\href{http://ascl.net/#1}{\nolinkurl{http://ascl.net/#1}}}
\providecommand{\doarXiv}[1]{\href{https://arxiv.org/abs/#1}{\nolinkurl{https://arxiv.org/abs/#1}}}

\bibitem[{M.~A. {Abramowicz} \& P.~C. {Fragile}(2013){Abramowicz} \&
  {Fragile}}]{2013Abramowicz}
{Abramowicz}, M.~A., \& {Fragile}, P.~C. 2013, \bibinfo{title}{{Foundations of
  Black Hole Accretion Disk Theory},} Living Reviews in Relativity, 16, 1,
  \dodoi{10.12942/lrr-2013-1}

\bibitem[{M.~A. {Ag{\"u}eros} {et~al.}(2009){Ag{\"u}eros}, {Anderson}, {Covey},
  {Hawley}, {Margon}, {Newsom}, {Posselt}, {Silvestri}, {Szkody}, \&
  {Voges}}]{2009ApJS..181..444A}
{Ag{\"u}eros}, M.~A., {Anderson}, S.~F., {Covey}, K.~R., {et~al.} 2009,
  \bibinfo{title}{{X-Ray-Emitting Stars Identified from the ROSAT All-Sky
  Survey and the Sloan Digital Sky Survey},} \apjs, 181, 444,
  \dodoi{10.1088/0067-0049/181/2/444}

\bibitem[{R. {Ahumada} {et~al.}(2020){Ahumada}, {Allende Prieto}, {Almeida},
  {Anders}, {Anderson}, {Andrews}, {Anguiano}, {Arcodia}, {Armengaud},
  {Aubert}, {Avila}, {Avila-Reese}, {Badenes}, {Balland}, {Barger},
  {Barrera-Ballesteros}, {Basu}, {Bautista}, {Beaton}, {Beers}, {Benavides},
  {Bender}, {Bernardi}, {Bershady}, {Beutler}, {Bidin}, {Bird}, {Bizyaev},
  {Blanc}, {Blanton}, {Boquien}, {Borissova}, {Bovy}, {Brandt}, {Brinkmann},
  {Brownstein}, {Bundy}, {Bureau}, {Burgasser}, {Burtin}, {Cano-D{\'\i}az},
  {Capasso}, {Cappellari}, {Carrera}, {Chabanier}, {Chaplin}, {Chapman},
  {Cherinka}, {Chiappini}, {Doohyun Choi}, {Chojnowski}, {Chung}, {Clerc},
  {Coffey}, {Comerford}, {Comparat}, {da Costa}, {Cousinou}, {Covey}, {Crane},
  {Cunha}, {Ilha}, {Dai}, {Damsted}, {Darling}, {Davidson}, {Davies}, {Dawson},
  {De}, {de la Macorra}, {De Lee}, {Queiroz}, {Deconto Machado}, {de la Torre},
  {Dell'Agli}, {du Mas des Bourboux}, {Diamond-Stanic}, {Dillon}, {Donor},
  {Drory}, {Duckworth}, {Dwelly}, {Ebelke}, {Eftekharzadeh}, {Davis Eigenbrot},
  {Elsworth}, {Eracleous}, {Erfanianfar}, {Escoffier}, {Fan}, {Farr},
  {Fern{\'a}ndez-Trincado}, {Feuillet}, {Finoguenov}, {Fofie},
  {Fraser-McKelvie}, {Frinchaboy}, {Fromenteau}, {Fu}, {Galbany}, {Garcia},
  {Garc{\'\i}a-Hern{\'a}ndez}, {Garma Oehmichen}, {Ge}, {Geimba Maia},
  {Geisler}, {Gelfand}, {Goddy}, {Gonzalez-Perez}, {Grabowski}, {Green},
  {Grier}, {Guo}, {Guy}, {Harding}, {Hasselquist}, {Hawken}, {Hayes}, {Hearty},
  {Hekker}, {Hogg}, {Holtzman}, {Horta}, {Hou}, {Hsieh}, {Huber}, {Hunt}, {Ider
  Chitham}, {Imig}, {Jaber}, {Jimenez Angel}, {Johnson}, {Jones},
  {J{\"o}nsson}, {Jullo}, {Kim}, {Kinemuchi}, {Kirkpatrick}, {Kite}, {Klaene},
  {Kneib}, {Kollmeier}, {Kong}, {Kounkel}, {Krishnarao}, {Lacerna}, {Lan},
  {Lane}, {Law}, {Le Goff}, {Leung}, {Lewis}, {Li}, {Lian}, {Lin}, {Long},
  {Longa-Pe{\~n}a}, {Lundgren}, {Lyke}, {Mackereth}, {MacLeod}, {Majewski},
  {Manchado}, {Maraston}, {Martini}, {Masseron}, {Masters}, {Mathur},
  {McDermid}, {Merloni}, {Merrifield}, {M{\'e}sz{\'a}ros}, {Miglio}, {Minniti},
  {Minsley}, {Miyaji}, {Mohammad}, {Mosser}, {Mueller}, {Muna},
  {Mu{\~n}oz-Guti{\'e}rrez}, {Myers}, {Nadathur}, {Nair}, {Nandra}, {Correa do
  Nascimento}, {Nevin}, {Newman}, {Nidever}, {Nitschelm}, {Noterdaeme},
  {O'Connell}, {Olmstead}, {Oravetz}, {Oravetz}, {Osorio}, {Pace}, {Padilla},
  {Palanque-Delabrouille}, \& {Palicio}}]{2020ApJS..249....3A}
{Ahumada}, R., {Allende Prieto}, C., {Almeida}, A., {et~al.} 2020,
  \bibinfo{title}{{The 16th Data Release of the Sloan Digital Sky Surveys:
  First Release from the APOGEE-2 Southern Survey and Full Release of eBOSS
  Spectra},} \apjs, 249, 3, \dodoi{10.3847/1538-4365/ab929e}

\bibitem[{A.~M. {Archibald} {et~al.}(2009){Archibald}, {Stairs}, {Ransom},
  {Kaspi}, {Kondratiev}, {Lorimer}, {McLaughlin}, {Boyles}, {Hessels}, {Lynch},
  {van Leeuwen}, {Roberts}, {Jenet}, {Champion}, {Rosen}, {Barlow}, {Dunlap},
  \& {Remillard}}]{2009Sci...324.1411A}
{Archibald}, A.~M., {Stairs}, I.~H., {Ransom}, S.~M., {et~al.} 2009,
  \bibinfo{title}{{A Radio Pulsar/X-ray Binary Link},} Science, 324, 1411,
  \dodoi{10.1126/science.1172740}

\bibitem[{A. {Bahramian} {et~al.}(2021){Bahramian}, {Heinke}, {Kennea},
  {Maccarone}, {Evans}, {Wijnands}, {Degenaar}, {in't Zand}, {Shaw}, {Rivera
  Sandoval}, {McClure}, {Tetarenko}, {Strader}, {Kuulkers}, \&
  {Sivakoff}}]{2021MNRAS.501.2790B}
{Bahramian}, A., {Heinke}, C.~O., {Kennea}, J.~A., {et~al.} 2021,
  \bibinfo{title}{{The Swift bulge survey: motivation, strategy, and first
  X-ray results},} \mnras, 501, 2790, \dodoi{10.1093/mnras/staa3868}

\bibitem[{E.~C. {Bellm} {et~al.}(2019){Bellm}, {Kulkarni}, {Graham}, {Dekany},
  {Smith}, {Riddle}, {Masci}, {Helou}, {Prince}, {Adams}, {Barbarino},
  {Barlow}, {Bauer}, {Beck}, {Belicki}, {Biswas}, {Blagorodnova}, {Bodewits},
  {Bolin}, {Brinnel}, {Brooke}, {Bue}, {Bulla}, {Burruss}, {Cenko}, {Chang},
  {Connolly}, {Coughlin}, {Cromer}, {Cunningham}, {De}, {Delacroix}, {Desai},
  {Duev}, {Eadie}, {Farnham}, {Feeney}, {Feindt}, {Flynn}, {Franckowiak},
  {Frederick}, {Fremling}, {Gal-Yam}, {Gezari}, {Giomi}, {Goldstein},
  {Golkhou}, {Goobar}, {Groom}, {Hacopians}, {Hale}, {Henning}, {Ho}, {Hover},
  {Howell}, {Hung}, {Huppenkothen}, {Imel}, {Ip}, {Ivezi{\'c}}, {Jackson},
  {Jones}, {Juric}, {Kasliwal}, {Kaspi}, {Kaye}, {Kelley}, {Kowalski},
  {Kramer}, {Kupfer}, {Landry}, {Laher}, {Lee}, {Lin}, {Lin}, {Lunnan},
  {Giomi}, {Mahabal}, {Mao}, {Miller}, {Monkewitz}, {Murphy}, {Ngeow},
  {Nordin}, {Nugent}, {Ofek}, {Patterson}, {Penprase}, {Porter}, {Rauch},
  {Rebbapragada}, {Reiley}, {Rigault}, {Rodriguez}, {van Roestel}, {Rusholme},
  {van Santen}, {Schulze}, {Shupe}, {Singer}, {Soumagnac}, {Stein}, {Surace},
  {Sollerman}, {Szkody}, {Taddia}, {Terek}, {Van Sistine}, {van Velzen},
  {Vestrand}, {Walters}, {Ward}, {Ye}, {Yu}, {Yan}, \&
  {Zolkower}}]{2019PASP..131a8002B}
{Bellm}, E.~C., {Kulkarni}, S.~R., {Graham}, M.~J., {et~al.} 2019,
  \bibinfo{title}{{The Zwicky Transient Facility: System Overview, Performance,
  and First Results},} \pasp, 131, 018002, \dodoi{10.1088/1538-3873/aaecbe}

\bibitem[{D. {Belloni} \& M.~R. {Schreiber}(2023){Belloni} \&
  {Schreiber}}]{2023hxga.book..129B}
{Belloni}, D., \& {Schreiber}, M.~R. 2023, in Handbook of X-ray and Gamma-ray
  Astrophysics, 129, \dodoi{10.1007/978-981-16-4544-0_98-1}

\bibitem[{T. {Boller} {et~al.}(2016){Boller}, {Freyberg}, {Tr{\"u}mper},
  {Haberl}, {Voges}, \& {Nandra}}]{2016A&A...588A.103B}
{Boller}, T., {Freyberg}, M.~J., {Tr{\"u}mper}, J., {et~al.} 2016,
  \bibinfo{title}{{Second ROSAT all-sky survey (2RXS) source catalogue},} \aap,
  588, A103, \dodoi{10.1051/0004-6361/201525648}

\bibitem[{J.~M. {Corral-Santana} {et~al.}(2016){Corral-Santana}, {Casares},
  {Mu{\~n}oz-Darias}, {Bauer}, {Mart{\'\i}nez-Pais}, \&
  {Russell}}]{2016A&A...587A..61C}
{Corral-Santana}, J.~M., {Casares}, J., {Mu{\~n}oz-Darias}, T., {et~al.} 2016,
  \bibinfo{title}{{BlackCAT: A catalogue of stellar-mass black holes in X-ray
  transients},} \aap, 587, A61, \dodoi{10.1051/0004-6361/201527130}

\bibitem[{X.-Q. {Cui} {et~al.}(2012){Cui}, {Zhao}, {Chu}, {Li}, {Li}, {Zhang},
  {Su}, {Yao}, {Wang}, {Xing}, {Li}, {Zhu}, {Wang}, {Gu}, {Luo}, {Xu}, {Zhang},
  {Liu}, {Zhang}, {Yang}, {Cao}, {Chen}, {Chen}, {Chen}, {Chen}, {Chu}, {Feng},
  {Gong}, {Hou}, {Hu}, {Hu}, {Hu}, {Jia}, {Jiang}, {Jiang}, {Jiang}, {Jin},
  {Li}, {Li}, {Li}, {Liu}, {Liu}, {Lu}, {Mao}, {Men}, {Qi}, {Qi}, {Shi},
  {Tang}, {Tao}, {Wang}, {Wang}, {Wang}, {Wang}, {Wang}, {Wang}, {Wang},
  {Wang}, {Wang}, {Wang}, {Wang}, {Wang}, {Xu}, {Xu}, {Yang}, {Yu}, {Yuan},
  {Yuan}, {Zhai}, {Zhang}, {Zhang}, {Zhang}, {Zhao}, {Zhou}, {Zhou}, {Zhu}, \&
  {Zou}}]{2012RAA....12.1197C}
{Cui}, X.-Q., {Zhao}, Y.-H., {Chu}, Y.-Q., {et~al.} 2012, \bibinfo{title}{{The
  Large Sky Area Multi-Object Fiber Spectroscopic Telescope (LAMOST)},}
  Research in Astronomy and Astrophysics, 12, 1197,
  \dodoi{10.1088/1674-4527/12/9/003}

\bibitem[{R.~A. {Downes} {et~al.}(2001){Downes}, {Webbink}, {Shara}, {Ritter},
  {Kolb}, \& {Duerbeck}}]{2001PASP..113..764D}
{Downes}, R.~A., {Webbink}, R.~F., {Shara}, M.~M., {et~al.} 2001,
  \bibinfo{title}{{A Catalog and Atlas of Cataclysmic Variables: The Living
  Edition},} \pasp, 113, 764, \dodoi{10.1086/320802}

\bibitem[{E.~D. {Feigelson} {et~al.}(2002){Feigelson}, {Broos}, {Gaffney},
  {Garmire}, {Hillenbrand}, {Pravdo}, {Townsley}, \&
  {Tsuboi}}]{2002ApJ...574..258F}
{Feigelson}, E.~D., {Broos}, P., {Gaffney}, III, J.~A., {et~al.} 2002,
  \bibinfo{title}{{X-Ray-emitting Young Stars in the Orion Nebula},} \apj, 574,
  258, \dodoi{10.1086/340936}

\bibitem[{F. {Fortin} {et~al.}(2024){Fortin}, {Kalsi}, {Garc{\'\i}a},
  {Simaz-Bunzel}, \& {Chaty}}]{2024A&A...684A.124F}
{Fortin}, F., {Kalsi}, A., {Garc{\'\i}a}, F., {Simaz-Bunzel}, A., \& {Chaty},
  S. 2024, \bibinfo{title}{{A catalogue of low-mass X-ray binaries in the
  Galaxy: From the INTEGRAL to the Gaia era},} \aap, 684, A124,
  \dodoi{10.1051/0004-6361/202347908}

\bibitem[{ {Gaia Collaboration} {et~al.}(2023){Gaia Collaboration},
  {Vallenari}, {Brown}, {Prusti}, {de Bruijne}, {Arenou}, {Babusiaux},
  {Biermann}, {Creevey}, {Ducourant}, {Evans}, {Eyer}, {Guerra}, {Hutton},
  {Jordi}, {Klioner}, {Lammers}, {Lindegren}, {Luri}, {Mignard}, {Panem},
  {Pourbaix}, {Randich}, {Sartoretti}, {Soubiran}, {Tanga}, {Walton},
  {Bailer-Jones}, {Bastian}, {Drimmel}, {Jansen}, {Katz}, {Lattanzi}, {van
  Leeuwen}, {Bakker}, {Cacciari}, {Casta{\~n}eda}, {De Angeli}, {Fabricius},
  {Fouesneau}, {Fr{\'e}mat}, {Galluccio}, {Guerrier}, {Heiter}, {Masana},
  {Messineo}, {Mowlavi}, {Nicolas}, {Nienartowicz}, {Pailler}, {Panuzzo},
  {Riclet}, {Roux}, {Seabroke}, {Sordo}, {Th{\'e}venin}, {Gracia-Abril},
  {Portell}, {Teyssier}, {Altmann}, {Andrae}, {Audard}, {Bellas-Velidis},
  {Benson}, {Berthier}, {Blomme}, {Burgess}, {Busonero}, {Busso},
  {C{\'a}novas}, {Carry}, {Cellino}, {Cheek}, {Clementini}, {Damerdji},
  {Davidson}, {de Teodoro}, {Nu{\~n}ez Campos}, {Delchambre}, {Dell'Oro},
  {Esquej}, {Fern{\'a}ndez-Hern{\'a}ndez}, {Fraile}, {Garabato},
  {Garc{\'\i}a-Lario}, {Gosset}, {Haigron}, {Halbwachs}, {Hambly}, {Harrison},
  {Hern{\'a}ndez}, {Hestroffer}, {Hodgkin}, {Holl}, {Jan{\ss}en}, {Jevardat de
  Fombelle}, {Jordan}, {Krone-Martins}, {Lanzafame}, {L{\"o}ffler}, {Marchal},
  {Marrese}, {Moitinho}, {Muinonen}, {Osborne}, {Pancino}, {Pauwels},
  {Recio-Blanco}, {Reyl{\'e}}, {Riello}, {Rimoldini}, {Roegiers}, {Rybizki},
  {Sarro}, {Siopis}, {Smith}, {Sozzetti}, {Utrilla}, {van Leeuwen}, {Abbas},
  {{\'A}brah{\'a}m}, {Abreu Aramburu}, {Aerts}, {Aguado}, {Ajaj},
  {Aldea-Montero}, {Altavilla}, {{\'A}lvarez}, {Alves}, {Anders}, {Anderson},
  {Anglada Varela}, {Antoja}, {Baines}, {Baker}, {Balaguer-N{\'u}{\~n}ez},
  {Balbinot}, {Balog}, {Barache}, {Barbato}, {Barros}, {Barstow},
  {Bartolom{\'e}}, {Bassilana}, {Bauchet}, {Becciani}, {Bellazzini},
  {Berihuete}, {Bernet}, {Bertone}, {Bianchi}, {Binnenfeld}, {Blanco-Cuaresma},
  {Blazere}, {Boch}, {Bombrun}, {Bossini}, {Bouquillon}, {Bragaglia},
  {Bramante}, {Breedt}, {Bressan}, {Brouillet}, {Brugaletta}, {Bucciarelli},
  {Burlacu}, {Butkevich}, {Buzzi}, {Caffau}, {Cancelliere}, {Cantat-Gaudin},
  {Carballo}, {Carlucci}, {Carnerero}, {Carrasco}, {Casamiquela}, {Castellani},
  {Castro-Ginard}, {Chaoul}, {Charlot}, {Chemin}, {Chiaramida}, {Chiavassa},
  {Chornay}, {Comoretto}, {Contursi}, {Cooper}, {Cornez}, {Cowell}, {Crifo},
  {Cropper}, {Crosta}, {Crowley}, {Dafonte}, {Dapergolas}, {David}, {David},
  {de Laverny}, {De Luise}, {De March}, {De Ridder}, {de Souza}, {de Torres},
  {del Peloso}, {del Pozo}, {Delbo}, {Delgado}, {Delisle}, {Demouchy},
  {Dharmawardena}, {Di Matteo}, {Diakite}, {Diener}, {Distefano}, {Dolding},
  {Edvardsson}, {Enke}, {Fabre}, {Fabrizio}, {Faigler}, {Fedorets}, {Fernique},
  {Fienga}, {Figueras}, {Fournier}, {Fouron}, {Fragkoudi}, {Gai},
  {Garcia-Gutierrez}, {Garcia-Reinaldos}, {Garc{\'\i}a-Torres}, {Garofalo},
  {Gavel}, {Gavras}, {Gerlach}, {Geyer}, {Giacobbe}, {Gilmore}, {Girona},
  {Giuffrida}, {Gomel}, {Gomez}, {Gonz{\'a}lez-N{\'u}{\~n}ez},
  {Gonz{\'a}lez-Santamar{\'\i}a}, {Gonz{\'a}lez-Vidal}, {Granvik}, {Guillout},
  {Guiraud}, {Guti{\'e}rrez-S{\'a}nchez}, {Guy}, {Hatzidimitriou}, {Hauser},
  {Haywood}, {Helmer}, {Helmi}, {Sarmiento}, {Hidalgo}, {Hilger},
  {H{\l}adczuk}, {Hobbs}, {Holland}, {Huckle}, {Jardine}, {Jasniewicz},
  {Jean-Antoine Piccolo}, {Jim{\'e}nez-Arranz}, {Jorissen}, {Juaristi
  Campillo}, {Julbe}, {Karbevska}, {Kervella}, {Khanna}, {Kontizas},
  {Kordopatis}, {Korn}, {K{\'o}sp{\'a}l}, {Kostrzewa-Rutkowska},
  {Kruszy{\'n}ska}, {Kun}, {Laizeau}, {Lambert}, {Lanza}, {Lasne}, {Le
  Campion}, {Lebreton}, {Lebzelter}, {Leccia}, {Leclerc}, {Lecoeur-Taibi},
  {Liao}, {Licata}, {Lindstr{\o}m}, {Lister}, {Livanou}, {Lobel}, {Lorca},
  {Loup}, {Madrero Pardo}, {Magdaleno Romeo}, {Managau}, {Mann}, {Manteiga},
  {Marchant}, {Marconi}, {Marcos}, {Marcos Santos}, {Mar{\'\i}n Pina},
  {Marinoni}, {Marocco}, {Marshall}, {Martin Polo}, {Mart{\'\i}n-Fleitas},
  {Marton}, {Mary}, {Masip}, {Massari}, {Mastrobuono-Battisti}, {Mazeh},
  {McMillan}, {Messina}, {Michalik}, {Millar}, {Mints}, {Molina}, {Molinaro},
  {Moln{\'a}r}, {Monari}, {Mongui{\'o}}, {Montegriffo}, {Montero}, {Mor},
  {Mora}, {Morbidelli}, {Morel}, {Morris}, {Muraveva}, {Murphy}, {Musella},
  {Nagy}, {Noval}, {Oca{\~n}a}, {Ogden}, {Ordenovic}, {Osinde}, {Pagani},
  {Pagano}, {Palaversa}, {Palicio}, {Pallas-Quintela}, {Panahi},
  {Payne-Wardenaar}, {Pe{\~n}alosa Esteller}, {Penttil{\"a}}, {Pichon},
  {Piersimoni}, {Pineau}, {Plachy}, {Plum}, {Poggio}, {Pr{\v{s}}a}, {Pulone},
  {Racero}, {Ragaini}, {Rainer}, {Raiteri}, {Rambaux}, {Ramos}, {Ramos-Lerate},
  {Re Fiorentin}, {Regibo}, {Richards}, {Rios Diaz}, {Ripepi}, {Riva}, {Rix},
  {Rixon}, {Robichon}, {Robin}, {Robin}, {Roelens}, {Rogues}, {Rohrbasser},
  {Romero-G{\'o}mez}, {Rowell}, {Royer}, {Ruz Mieres}, {Rybicki}, {Sadowski},
  {S{\'a}ez N{\'u}{\~n}ez}, {Sagrist{\`a} Sell{\'e}s}, {Sahlmann}, {Salguero},
  {Samaras}, {Sanchez Gimenez}, {Sanna}, {Santove{\~n}a}, {Sarasso},
  {Schultheis}, {Sciacca}, {Segol}, {Segovia}, {S{\'e}gransan}, {Semeux},
  {Shahaf}, {Siddiqui}, {Siebert}, {Siltala}, {Silvelo}, {Slezak}, {Slezak},
  {Smart}, {Snaith}, {Solano}, {Solitro}, {Souami}, {Souchay}, {Spagna},
  {Spina}, {Spoto}, {Steele}, {Steidelm{\"u}ller}, {Stephenson}, {S{\"u}veges},
  {Surdej}, {Szabados}, {Szegedi-Elek}, {Taris}, {Taylor}, {Teixeira},
  {Tolomei}, {Tonello}, {Torra}, {Torra}, {Torralba Elipe}, {Trabucchi},
  {Tsounis}, {Turon}, {Ulla}, {Unger}, {Vaillant}, {van Dillen}, {van Reeven},
  {Vanel}, {Vecchiato}, {Viala}, {Vicente}, {Voutsinas}, {Weiler}, {Wevers},
  {Wyrzykowski}, {Yoldas}, {Yvard}, {Zhao}, {Zorec}, {Zucker}, \&
  {Zwitter}}]{GaiaDR3}
{Gaia Collaboration}, {Vallenari}, A., {Brown}, A.~G.~A., {et~al.} 2023,
  \bibinfo{title}{{Gaia Data Release 3. Summary of the content and survey
  properties},} \aap, 674, A1, \dodoi{10.1051/0004-6361/202243940}

\bibitem[{C.~J. {Hailey} {et~al.}(2018){Hailey}, {Mori}, {Bauer}, {Berkowitz},
  {Hong}, \& {Hord}}]{2018Natur.556...70H}
{Hailey}, C.~J., {Mori}, K., {Bauer}, F.~E., {et~al.} 2018, \bibinfo{title}{{A
  density cusp of quiescent X-ray binaries in the central parsec of the
  Galaxy},} \nat, 556, 70, \dodoi{10.1038/nature25029}

\bibitem[{C.~O. {Heinke} {et~al.}(2003){Heinke}, {Grindlay}, {Lugger}, {Cohn},
  {Edmonds}, {Lloyd}, \& {Cool}}]{2003ApJ...598..501H}
{Heinke}, C.~O., {Grindlay}, J.~E., {Lugger}, P.~M., {et~al.} 2003,
  \bibinfo{title}{{Analysis of the Quiescent Low-Mass X-Ray Binary Population
  in Galactic Globular Clusters},} \apj, 598, 501, \dodoi{10.1086/378885}

\bibitem[{D.~J. {Helfand}(1980){Helfand}}]{1980PASP...92..691H}
{Helfand}, D.~J. 1980, \bibinfo{title}{{A search for X-ray binary stars in
  their quiescent phase.},} \pasp, 92, 691, \dodoi{10.1086/130731}

\bibitem[{R.~I. {Hynes} {et~al.}(2011){Hynes}, {Jonker}, {Bassa}, {Dieball},
  {Greiss}, {Maccarone}, {Nelemans}, {Steeghs}, {Torres}, {Britt}, {Clem},
  {Gossen}, {Grindlay}, {Groot}, {Kuiper}, {Kuulkers}, {Mendez}, {Mikles},
  {Ratti}, {Rea}, {van Haaften}, {Wijnands}, \& {in't
  Zand}}]{2011AAS...21714424H}
{Hynes}, R.~I., {Jonker}, P.~G., {Bassa}, C.~G., {et~al.} 2011, in American
  Astronomical Society Meeting Abstracts, Vol. 217, American Astronomical
  Society Meeting Abstracts \#217, 144.24

\bibitem[{V.~M. {Kaspi} \& A.~M. {Beloborodov}(2017){Kaspi} \&
  {Beloborodov}}]{2017ARA&A..55..261K}
{Kaspi}, V.~M., \& {Beloborodov}, A.~M. 2017, \bibinfo{title}{{Magnetars},}
  \araa, 55, 261, \dodoi{10.1146/annurev-astro-081915-023329}

\bibitem[{U. {Kolb}(1998){Kolb}}]{1998ASPC..137..190K}
{Kolb}, U. 1998, in Astronomical Society of the Pacific Conference Series, Vol.
  137, Wild Stars in the Old West, ed. S.~{Howell}, E.~{Kuulkers}, \&
  C.~{Woodward}, 190, \dodoi{10.48550/arXiv.astro-ph/9708188}

\bibitem[{J.~P. {Lasota}(2000){Lasota}}]{2000A&A...360..575L}
{Lasota}, J.~P. 2000, \bibinfo{title}{{X-rays from quiescent low-mass X-ray
  binary transients},} \aap, 360, 575, \dodoi{10.48550/arXiv.astro-ph/0005073}

\bibitem[{A. {Liu} {et~al.}(2022){Liu}, {Bulbul}, {Ghirardini}, {Liu}, {Klein},
  {Clerc}, {{\"O}zsoy}, {Ramos-Ceja}, {Pacaud}, {Comparat}, {Okabe}, {Bahar},
  {Biffi}, {Brunner}, {Br{\"u}ggen}, {Buchner}, {Ider Chitham}, {Chiu},
  {Dolag}, {Gatuzz}, {Gonzalez}, {Hoang}, {Lamer}, {Merloni}, {Nandra},
  {Oguri}, {Ota}, {Predehl}, {Reiprich}, {Salvato}, {Schrabback}, {Sanders},
  {Seppi}, \& {Thibaud}}]{2022A&A...661A...2L}
{Liu}, A., {Bulbul}, E., {Ghirardini}, V., {et~al.} 2022, \bibinfo{title}{{The
  eROSITA Final Equatorial-Depth Survey (eFEDS). Catalog of galaxy clusters and
  groups},} \aap, 661, A2, \dodoi{10.1051/0004-6361/202141120}

\bibitem[{Q.~Z. {Liu} {et~al.}(2007){Liu}, {van Paradijs}, \& {van den
  Heuvel}}]{2007A&A...469..807L}
{Liu}, Q.~Z., {van Paradijs}, J., \& {van den Heuvel}, E.~P.~J. 2007,
  \bibinfo{title}{{A catalogue of low-mass X-ray binaries in the Galaxy, LMC,
  and SMC (Fourth edition)},} \aap, 469, 807,
  \dodoi{10.1051/0004-6361:20077303}

\bibitem[{N.~R. {Lomb}(1976){Lomb}}]{1976Ap&SS..39..447L}
{Lomb}, N.~R. 1976, \bibinfo{title}{{Least-Squares Frequency Analysis of
  Unequally Spaced Data},} \apss, 39, 447, \dodoi{10.1007/BF00648343}

\bibitem[{A. {Merloni} {et~al.}(2024){Merloni}, {Lamer}, {Liu}, {Ramos-Ceja},
  {Brunner}, {Bulbul}, {Dennerl}, {Doroshenko}, {Freyberg}, {Friedrich},
  {Gatuzz}, {Georgakakis}, {Haberl}, {Igo}, {Kreykenbohm}, {Liu}, {Maitra},
  {Malyali}, {Mayer}, {Nandra}, {Predehl}, {Robrade}, {Salvato}, {Sanders},
  {Stewart}, {Tub{\'\i}n-Arenas}, {Weber}, {Wilms}, {Arcodia}, {Artis},
  {Aschersleben}, {Avakyan}, {Aydar}, {Bahar}, {Balzer}, {Becker}, {Berger},
  {Boller}, {Bornemann}, {Br{\"u}ggen}, {Brusa}, {Buchner}, {Burwitz},
  {Camilloni}, {Clerc}, {Comparat}, {Coutinho}, {Czesla}, {Dannhauer},
  {Dauner}, {Dauser}, {Dietl}, {Dolag}, {Dwelly}, {Egg}, {Ehl}, {Freund},
  {Friedrich}, {Gaida}, {Garrel}, {Ghirardini}, {Gokus}, {Gr{\"u}nwald},
  {Grandis}, {Grotova}, {Gruen}, {Gueguen}, {H{\"a}mmerich}, {Hamaus},
  {Hasinger}, {Haubner}, {Homan}, {Ider Chitham}, {Joseph}, {Joyce},
  {K{\"o}nig}, {Kaltenbrunner}, {Khokhriakova}, {Kink}, {Kirsch}, {Kluge},
  {Knies}, {Krippendorf}, {Krumpe}, {Kurpas}, {Li}, {Liu}, {Locatelli},
  {Lorenz}, {M{\"u}ller}, {Magaudda}, {Mannes}, {McCall}, {Meidinger},
  {Michailidis}, {Migkas}, {Mu{\~n}oz-Giraldo}, {Musiimenta}, {Nguyen-Dang},
  {Ni}, {Olechowska}, {Ota}, {Pacaud}, {Pasini}, {Perinati}, {Pires},
  {Pommranz}, {Ponti}, {Poppenhaeger}, {P{\"u}hlhofer}, {Rau}, {Reh},
  {Reiprich}, {Roster}, {Saeedi}, {Santangelo}, {Sasaki}, {Schmitt},
  {Schneider}, {Schrabback}, {Schuster}, {Schwope}, {Seppi}, {Serim},
  {Shreeram}, {Sokolova-Lapa}, {Starck}, {Stelzer}, {Stierhof}, {Suleimanov},
  {Tenzer}, {Traulsen}, {Tr{\"u}mper}, {Tsuge}, {Urrutia}, {Veronica},
  {Waddell}, {Willer}, {Wolf}, {Yeung}, {Zainab}, {Zangrandi}, {Zhang},
  {Zhang}, \& {Zheng}}]{2024A&A...682A..34M}
{Merloni}, A., {Lamer}, G., {Liu}, T., {et~al.} 2024, \bibinfo{title}{{The
  SRG/eROSITA all-sky survey. First X-ray catalogues and data release of the
  western Galactic hemisphere},} \aap, 682, A34,
  \dodoi{10.1051/0004-6361/202347165}

\bibitem[{C. {Motch} {et~al.}(1998){Motch}, {Guillout}, {Haberl}, {Krautter},
  {Pakull}, {Pietsch}, {Reinsch}, {Voges}, \& {Zickgraf}}]{1998A&AS..132..341M}
{Motch}, C., {Guillout}, P., {Haberl}, F., {et~al.} 1998,
  \bibinfo{title}{{Identification of selected sources from the ROSAT Galactic
  Plane Survey. I.},} \aaps, 132, 341, \dodoi{10.1051/aas:1998299}

\bibitem[{A.~F. {Pala} {et~al.}(2020){Pala}, {G{\"a}nsicke}, {Breedt},
  {Knigge}, {Hermes}, {Gentile Fusillo}, {Hollands}, {Naylor}, {Pelisoli},
  {Schreiber}, {Toonen}, {Aungwerojwit}, {Cukanovaite}, {Dennihy}, {Manser},
  {Pretorius}, {Scaringi}, \& {Toloza}}]{2020MNRAS.494.3799P}
{Pala}, A.~F., {G{\"a}nsicke}, B.~T., {Breedt}, E., {et~al.} 2020,
  \bibinfo{title}{{A Volume-limited Sample of Cataclysmic Variables from Gaia
  DR2: Space Density and Population Properties},} \mnras, 494, 3799,
  \dodoi{10.1093/mnras/staa764}

\bibitem[{K.~A. {Postnov} \& L.~R. {Yungelson}(2014){Postnov} \&
  {Yungelson}}]{2014LRR....17....3P}
{Postnov}, K.~A., \& {Yungelson}, L.~R. 2014, \bibinfo{title}{{The Evolution of
  Compact Binary Star Systems},} Living Reviews in Relativity, 17, 3,
  \dodoi{10.12942/lrr-2014-3}

\bibitem[{P. {Predehl} {et~al.}(2021){Predehl}, {Andritschke}, {Arefiev},
  {Babyshkin}, {Batanov}, {Becker}, {B{\"o}hringer}, {Bogomolov}, {Boller},
  {Borm}, {Bornemann}, {Br{\"a}uninger}, {Br{\"u}ggen}, {Brunner}, {Brusa},
  {Bulbul}, {Buntov}, {Burwitz}, {Burkert}, {Clerc}, {Churazov}, {Coutinho},
  {Dauser}, {Dennerl}, {Doroshenko}, {Eder}, {Emberger}, {Eraerds},
  {Finoguenov}, {Freyberg}, {Friedrich}, {Friedrich}, {F{\"u}rmetz},
  {Georgakakis}, {Gilfanov}, {Granato}, {Grossberger}, {Gueguen}, {Gureev},
  {Haberl}, {H{\"a}lker}, {Hartner}, {Hasinger}, {Huber}, {Ji}, {Kienlin},
  {Kink}, {Korotkov}, {Kreykenbohm}, {Lamer}, {Lomakin}, {Lapshov}, {Liu},
  {Maitra}, {Meidinger}, {Menz}, {Merloni}, {Mernik}, {Mican}, {Mohr},
  {M{\"u}ller}, {Nandra}, {Nazarov}, {Pacaud}, {Pavlinsky}, {Perinati},
  {Pfeffermann}, {Pietschner}, {Ramos-Ceja}, {Rau}, {Reiffers}, {Reiprich},
  {Robrade}, {Salvato}, {Sanders}, {Santangelo}, {Sasaki}, {Scheuerle},
  {Schmid}, {Schmitt}, {Schwope}, {Shirshakov}, {Steinmetz}, {Stewart},
  {Str{\"u}der}, {Sunyaev}, {Tenzer}, {Tiedemann}, {Tr{\"u}mper}, {Voron},
  {Weber}, {Wilms}, \& {Yaroshenko}}]{Predehl2021}
{Predehl}, P., {Andritschke}, R., {Arefiev}, V., {et~al.} 2021,
  \bibinfo{title}{{The eROSITA X-ray telescope on SRG},} \aap, 647, A1,
  \dodoi{10.1051/0004-6361/202039313}

\bibitem[{R.~A. {Remillard} \& J.~E. {McClintock}(2006){Remillard} \&
  {McClintock}}]{2006Remillard}
{Remillard}, R.~A., \& {McClintock}, J.~E. 2006, \bibinfo{title}{{X-Ray
  Properties of Black-Hole Binaries},} \araa, 44, 49,
  \dodoi{10.1146/annurev.astro.44.051905.092532}

\bibitem[{A.~C. {Rodriguez}(2024){Rodriguez}}]{Rodriguez2024}
{Rodriguez}, A.~C. 2024, \bibinfo{title}{{From Active Stars to Black Holes: A
  Discovery Tool for Galactic X-Ray Sources},} \pasp, 136, 054201,
  \dodoi{10.1088/1538-3873/ad357c}

\bibitem[{A.~C. {Rodriguez} {et~al.}(2025){Rodriguez}, {El-Badry},
  {Suleimanov}, {Pala}, {Kulkarni}, {Gaensicke}, {Mori}, {Rich}, {Sarkar},
  {Bao}, {Lopes de Oliveira}, {Ramsay}, {Szkody}, {Graham}, {Prince},
  {Caiazzo}, {Vanderbosch}, {van Roestel}, {Das}, {Qin}, {Kasliwal}, {Wold},
  {Groom}, {Reiley}, \& {Riddle}}]{2025PASP..137a4201R}
{Rodriguez}, A.~C., {El-Badry}, K., {Suleimanov}, V., {et~al.} 2025,
  \bibinfo{title}{{Cataclysmic Variables and AM CVn Binaries in SRG/eROSITA +
  Gaia: Volume Limited Samples, X-Ray Luminosity Functions, and Space
  Densities},} \pasp, 137, 014201, \dodoi{10.1088/1538-3873/ada185}

\bibitem[{S. {Saeedi} {et~al.}(2022){Saeedi}, {Liu}, {Knies}, {Sasaki},
  {Becker}, {Bulbul}, {Dennerl}, {Freyberg}, {Laktionov}, \&
  {Merloni}}]{2022A&A...661A..35S}
{Saeedi}, S., {Liu}, T., {Knies}, J., {et~al.} 2022, \bibinfo{title}{{eROSITA
  study of the globular cluster 47 Tucanae},} \aap, 661, A35,
  \dodoi{10.1051/0004-6361/202141612}

\bibitem[{M. {Salvato} {et~al.}(2022){Salvato}, {Wolf}, {Dwelly},
  {Georgakakis}, {Brusa}, {Merloni}, {Liu}, {Toba}, {Nandra}, {Lamer},
  {Buchner}, {Schneider}, {Freund}, {Rau}, {Schwope}, {Nishizawa}, {Klein},
  {Arcodia}, {Comparat}, {Musiimenta}, {Nagao}, {Brunner}, {Malyali},
  {Finoguenov}, {Anderson}, {Shen}, {Ibarra-Medel}, {Trump}, {Brandt}, {Urry},
  {Rivera}, {Krumpe}, {Urrutia}, {Miyaji}, {Ichikawa}, {Schneider}, {Fresco},
  {Boller}, {Haase}, {Brownstein}, {Lane}, {Bizyaev}, \&
  {Nitschelm}}]{2022A&A...661A...3S}
{Salvato}, M., {Wolf}, J., {Dwelly}, T., {et~al.} 2022, \bibinfo{title}{{The
  eROSITA Final Equatorial-Depth Survey (eFEDS). Identification and
  characterization of the counterparts to point-like sources},} \aap, 661, A3,
  \dodoi{10.1051/0004-6361/202141631}

\bibitem[{J.~D. {Scargle}(1981){Scargle}}]{1981ApJS...45....1S}
{Scargle}, J.~D. 1981, \bibinfo{title}{{Studies in astronomical time series
  analysis. I - Modeling random processes in the time domain},} \apjs, 45, 1,
  \dodoi{10.1086/190706}

\bibitem[{D.~A. {Smith} \& V.~S. {Dhillon}(1998){Smith} \&
  {Dhillon}}]{1998MNRAS.301..767S}
{Smith}, D.~A., \& {Dhillon}, V.~S. 1998, \bibinfo{title}{{The secondary stars
  in cataclysmic variables and low-mass X-ray binaries},} \mnras, 301, 767,
  \dodoi{10.1046/j.1365-8711.1998.02065.x}

\bibitem[{F. {Verbunt} {et~al.}(1997){Verbunt}, {Bunk}, {Ritter}, \&
  {Pfeffermann}}]{1997A&A...327..602V}
{Verbunt}, F., {Bunk}, W.~H., {Ritter}, H., \& {Pfeffermann}, E. 1997,
  \bibinfo{title}{{Cataclysmic variables in the ROSAT PSPC All Sky Survey.},}
  \aap, 327, 602

\bibitem[{D.~A. {Verner} {et~al.}(1996){Verner}, {Ferland}, {Korista}, \&
  {Yakovlev}}]{1996ApJ...465..487V}
{Verner}, D.~A., {Ferland}, G.~J., {Korista}, K.~T., \& {Yakovlev}, D.~G. 1996,
  \bibinfo{title}{{Atomic Data for Astrophysics. II. New Analytic Fits for
  Photoionization Cross Sections of Atoms and Ions},} \apj, 465, 487,
  \dodoi{10.1086/177435}

\bibitem[{W. {Voges} {et~al.}(1999){Voges}, {Aschenbach}, {Boller},
  {Br{\"a}uninger}, {Briel}, {Burkert}, {Dennerl}, {Englhauser}, {Gruber},
  {Haberl}, {Hartner}, {Hasinger}, {K{\"u}rster}, {Pfeffermann}, {Pietsch},
  {Predehl}, {Rosso}, {Schmitt}, {Tr{\"u}mper}, \&
  {Zimmermann}}]{1999A&A...349..389V}
{Voges}, W., {Aschenbach}, B., {Boller}, T., {et~al.} 1999,
  \bibinfo{title}{{The ROSAT all-sky survey bright source catalogue},} \aap,
  349, 389, \dodoi{10.48550/arXiv.astro-ph/9909315}

\bibitem[{X. {Wang} \& J. {Takata}(2025){Wang} \& {Takata}}]{Wang2024}
{Wang}, X., \& {Takata}, J. 2025, \bibinfo{title}{{Cataclysmic Variable
  Candidates Identified in eROSITA-DE DR1,XMM-Newton, Swift, and ROSAT
  Catalogs},} arXiv e-prints, arXiv:2504.10794,
  \dodoi{10.48550/arXiv.2504.10794}

\bibitem[{B. {Warner}(1995){Warner}}]{1995Warner}
{Warner}, B. 1995, {Cataclysmic variable stars}, Vol.~28

\bibitem[{J. {Wilms} {et~al.}(2000){Wilms}, {Allen}, \&
  {McCray}}]{2000ApJ...542..914W}
{Wilms}, J., {Allen}, A., \& {McCray}, R. 2000, \bibinfo{title}{{On the
  Absorption of X-Rays in the Interstellar Medium},} \apj, 542, 914,
  \dodoi{10.1086/317016}

\bibitem[{Z.-X. {Zhang} {et~al.}(2022){Zhang}, {Zheng}, {Gu}, {Sun}, {Yi},
  {Shi}, {Wang}, {Bai}, {Zhang}, {Cui}, {Wang}, {Wu}, {Li}, {Shao}, {Lu},
  {Bai}, {Li}, {Fu}, \& {Liu}}]{2022ApJ...933..193Z}
{Zhang}, Z.-X., {Zheng}, L.-L., {Gu}, W.-M., {et~al.} 2022, \bibinfo{title}{{A
  Long-period Pre-ELM System Discovered from the LAMOST Medium-resolution
  Survey},} \apj, 933, 193, \dodoi{10.3847/1538-4357/ac75b6}

\end{thebibliography}
\bibliographystyle{aasjournalv7}

\end{document}